\documentclass[twocolumn]{aastex701}

\def\MJ{M_\mathrm{J}}

\def\Teff{T_{\text{eff}}}

\usepackage{url}
\usepackage{amsmath}

\begin{document}

\title{Direct Measurement of Extinction in a Planet-Hosting Gap}

\author[0000-0001-7255-3251]{Gabriele Cugno}
\affiliation{Department of Astrophysics, University of Z\"urich, Winterthurerstrasse 190, 8057 Z\"urich, Switzerland}
\affiliation{Department of Astronomy, University of Michigan, 1085 S. University Ave., Ann Arbor, MI 48109, USA}
\email[show]{gabriele.cugno@uzh.ch}  
\author[0000-0003-4689-2684]{Stefano Facchini}
\affiliation{Dipartimento di Fisica, Università degli Studi di Milano, Via Celoria 16, 20133 Milano, Italy}
\email[]{} 
\author[0000-0002-2692-7862]{Felipe Alarcon}
\affiliation{Dipartimento di Fisica, Università degli Studi di Milano, Via Celoria 16, 20133 Milano, Italy}
\email[]{} 
\author[0000-0001-7258-770X]{Jaehan Bae}
\affiliation{Department of Astronomy, University of Florida, Gainesville, FL 32611, USA}
\email[]{} 
\author[0000-0002-7695-7605]{Myriam Benisty}
\affiliation{Max-Planck Institute for Astronomy (MPIA), Königstuhl 17, 69117 Heidelberg, Germany}
\email[]{} 
\author[0000-0003-2895-6218]{Anna-Christina Eilers}
\affiliation{MIT Kavli Institute for Astrophysics and Space Research, 77 Massachusetts Avenue, Cambridge, 02139, Massachusetts, USA}
\email[]{} 
\author[0000-0002-9393-6507]{Gene C. K. Leung}
\affiliation{MIT Kavli Institute for Astrophysics and Space Research, 77 Massachusetts Avenue, Cambridge, 02139, Massachusetts, USA}
\email[]{} 
\author[0000-0003-1227-3084]{Michael Meyer}
\affiliation{Department of Astronomy, University of Michigan, 1085 S. University Ave., Ann Arbor, MI 48109, USA}
\email[]{} 
\author{Laurent Pueyo}
\affiliation{Space Telescope Science Institute, 3700 San Martin Drive, Baltimore, MD 21218, USA}
\email[]{} 
\author[0000-0003-1534-5186]{Richard Teague}
\affiliation{Department of Earth, Atmospheric, and Planetary Sciences, Massachusetts Institute of Technology, Cambridge, MA 02139, USA}
\email[]{} 
\author[0000-0003-4179-6394]{Edwin Bergin}
\affiliation{Department of Astronomy, University of Michigan, 1085 S. University Ave., Ann Arbor, MI 48109, USA}
\email[]{} 
\author[0000-0001-8627-0404]{Julien Girard}
\affiliation{Space Telescope Science Institute, 3700 San Martin Drive, Baltimore, MD 21218, USA}
\email[]{} 
\author[0000-0001-5555-2652]{Ravit Helled}
\affiliation{Department of Astrophysics, University of Z\"urich, Winterthurerstrasse 190, 8057 Z\"urich, Switzerland}
\email[]{} 
\author[0000-0001-6947-6072]{Jane Huang}
\affiliation{Department of Astronomy, Columbia University, 538 W. 120th Street, Pupin Hall, New York, NY 10027, USA}
\email[]{} 
\author[0000-0002-0834-6140]{Jarron Leisenring}
\affiliation{Steward Observatory and Department of Astronomy, University of Arizona, 933 N Cherry Avenue, Tucson, AZ 85721, USA}
\email[]{}



\begin{abstract}

Recent disk observations have revealed multiple indirect signatures of forming gas giant planets, but high-contrast imaging has rarely confirmed the presence of the suspected perturbers. 
Here, we exploit a unique opportunity provided by the background star AS~209bkg, which shines through a wide annular gap in the AS~209 disk, to perform transmission spectrophotometry and directly measure the extinction from gap material for the first time. 
By combining new VLT/SPHERE and JWST/NIRCam observations with archival HST data from 2005, we model the spectral energy distribution (SED) of AS~209bkg over a 19-year baseline. We find that the SED and its variability are best explained by increasing extinction along the line of sight as AS~209bkg approaches the gap edge in projection. The extinction is best described by a combination of ISM-like extinction component and a grey extinction component. This points to the presence of grains in the disk outer gap that are larger than in the ISM.
We find that the extinction in the gap at $\lambda\sim4.0~\mu$m is $A_{4\,\mu\mathrm{m}} = 2.7^{+0.7}_{-0.7}$~mag, while at H$\alpha$ ($\lambda=0.656~\mu$m), where most searches for accretion signatures take place, the extinction could be as high as $A_\mathrm{H\alpha} = 4.2^{+0.9}_{-1.2}$~mag ($A_V=4.6^{+1.0}_{-1.3}$). This suggests that even wide, deep gaps can significantly obscure emission from protoplanets, even those following a hot-start evolutionary model. Our extinction measurements help reconcile the discrepancy between ALMA-based predictions of planet-disk interactions and the non-detections from sensitive optical and near-infrared imaging campaigns.

\end{abstract}

\keywords{Exoplanet formation (492) --- High contrast techniques(2369) --- Circumstellar disks(235)}


\section{Introduction} \label{sec:intro}
Over the past decade, high angular resolution observations of protoplanetary disks, supported by numerical simulations, have revealed a variety of structures that indirectly signal the presence of forming gas giant planets. These include rings, gaps, and spiral arms traced by millimeter-sized grains in the sub-millimeter \citep[e.g.,][]{Andrews2018}, and by micron-sized grains in the near-infrared \citep[NIR; e.g.,][]{Benisty2023}, as well as kinematic perturbations in the gas dynamics \citep{Pinte2023} and chemical anomalies \citep{Law2023}. These findings have fueled a major effort in the high-contrast imaging community to directly detect forming protoplanets. Most searches have concentrated on two spectral regimes: H$\alpha$ in the optical, tracing accretion, and the NIR, tracing the planetary thermal emission.

\begin{table*}[t!]
\centering
\caption{Summary of the VLT/SPHERE observations.}
\def\arraystretch{1.25}
\begin{tabular}{lllllllllllll}\hline
Instrument & Prog. ID & Obs date & Filter & Coronagraph & DIT & \# of & Field  & Airmass & Seeing  \\
& & (yyyy-mm-dd) & & & & DITs & Rot. ($^\circ$) & & ($^{\prime\prime}$) \\ \hline
IRDIS & 113.268U.001 & 2024-06-17 & B\_Y & N\_ALC\_YJH\_S & 32 & 32 & 68.1 & 1.01-1.06 & 0.96       \\
IRDIS & 113.268U.001 & 2024-06-08 & B\_J & N\_ALC\_YJH\_S & 32 & 32 & 62.6 & 1.01-1.06 & 0.52-0.8       \\\hline
\end{tabular}\\\vspace{0.2cm}
\label{tab:observations_sphere}
\end{table*}

Similar to their young parent star, H$\alpha$ emission is expected from gas accreting onto a forming planet \citep[e.g.,][]{Aoyama2018}. Multiple surveys using instruments such as MagAO, SPHERE/ZIMPOL, VAMPIRES, and MUSE have targeted this emission line \citep{Cugno2019, Zurlo2020, Xie2020, Huelamo2022, Follette2023}. However, robust detections remain rare. Only the PDS~70 system is confirmed to host two accreting protoplanets with detections of H$\alpha$ emission \citep{Keppler2018, Haffert2019, Zhou2021, Close2025, Zhou2025}. Other candidates, such as AB~Aur~b \citep{Currie2022} and HD169142~b \citep{Hammond2023}, remain unconfirmed \citep{Zhou2023, Bowler2025}.

In the infrared, surveys with VLT/SPHERE \citep{AsensioTorres2021, Ren2023}, VLT/NaCo \citep{Jorquera2021, Cugno2023_ISPY}, and Keck/NIRC2 \citep{Wallack2024} searched for thermal emission from contracting protoplanets. Some of these surveys target wavelengths around 4~$\mu$m, where scattered light from the disk is reduced and extinction is presumed to be less severe than at shorter wavelengths. Despite these efforts, detections remain elusive in this spectral range as well. Among the few candidates detections, some exhibit unusually red spectra that may hint at significant dust obscuration along the line of sight (e.g., MWC758c, \citealt{Wagner2023}). 

With its unprecedented sensitivity at large separations ($\gtrsim1\farcs0$), JWST was expected to uncover a missing population of forming gas giants on wide orbits. Yet, early high-contrast imaging campaigns with JWST/NIRCam targeting disks with strong signatures of planet-disk interaction have also yielded non-detections \citep{Cugno2024, Wagner2024, Mullin2024, Uyama2025}. For example, in SAO~206462, \citet{Cugno2024} ruled out planets more massive than $\sim$2~$M_\mathrm{J}$ assuming a hot-start model, despite hydrodynamical simulations \citep[e.g.,][]{FungDong2015, Bae2016, DongFung2017} predicting that a $\gtrsim5~M_\mathrm{J}$ perturber is required to launch the observed spiral arms \citep{Stolker2016}.

One proposed explanation is that extinction from both circumstellar and circumplanetary disk material obscures the emission from forming planets at these wavelengths \citep[e.g.,][]{Alarcon2024}. However, very little is known about the magnitude and wavelength dependence of extinction in planet-forming environments. Estimates for known systems vary widely: \citet{Hashimoto2020} constrained the extinction toward PDS~70~b to be $A_{H\alpha} \gtrsim 2$~mag using the combined information of H$\alpha$ detection and H$\beta$ non-detection, while \citet{Uyama2021} used a Pa$\beta$ non-detection to derive a lower value of $A_{H\alpha} \sim 0.7$~mag. Conversely, the candidate MWC~758~c may suffer from $A_V \gtrsim 50$~mag of ISM-like extinction if it follows a hot-start model \citep{Wagner2023}. 

\begin{table*}[t!]
\centering
\caption{Summary of the JWST/NIRCam observations.}
\def\arraystretch{1.25}
\begin{tabular}{lllllllllllll}\hline
Instrument  & Prog. ID  & Obs date      & Target    & Filter      & Coronagraph   & N$_\mathrm{gr}$ & N$_\mathrm{int}$ & N$_\mathrm{dither}$ & N$_\mathrm{roll}$ & Readout\\
            &           & (yyyy-mm-dd)  &           & (SW/LW)     &               &                 &                  &                     &                   &\\ \hline
NIRCam      & 2487      & 2022-08-10    & AS~209     & F200W/F410M & $-$           & 10              & 160              & 4                   & 2 & RAPID   \\
NIRCam      & 5816      & 2024-08-13    & HD153135  & F182M/F300M & M335R         & 7               & 3                & 5                   & 1 & MEDIUM2      \\
NIRCam      & 5816      & 2024-08-13    & HD153135  & F210M/F335M & M335R         & 7               & 3                & 5                   & 1 & MEDIUM2       \\
NIRCam      & 5816      & 2024-08-13    & AS~209     & F182M/F300M & M335R         & 19              & 5                & 1                   & 2 & DEEP8      \\
NIRCam      & 5816      & 2024-08-13    & AS~209     & F210M/F335M & M335R         & 19              & 5                & 1                   & 2 & DEEP8       \\
NIRCam      & 5816      & 2024-08-13    & AS~209     & F115W/F250M & $-$           & 10              & 40               & 4                   & 2 & RAPID       \\
NIRCam      & 5816      & 2024-08-13    & AS~209     & F090W/F323N & $-$           & 10              & 30               & 4                   & 2 & RAPID      \\
NIRCam      & 5816      & 2024-08-13    & AS~209     & F070W/F405N & $-$           & 10              & 30               & 4                   & 2 & RAPID       \\
\hline
\end{tabular}\\\vspace{0.2cm}
\label{tab:observations_nircam}
\end{table*}

Even for the continuum emission of the well-characterized PDS~70 planets, inferred extinction values depend strongly on the model assumptions. For instance, fits of the Spectral Energy Distributions (SEDs) yield $A_V$ values of $3-15$ mag \citep{Wang2021}, supported by the non-detection of atmospheric molecular features at medium spectral resolution \citep{Cugno2021}. Notably, PDS~70~b and c lie within a large cavity ($\sim$20–30~au), where small grain depletion is significant. In contrast, many other putative protoplanets are thought to reside in narrower dust gaps seen in ALMA continuum images. As shown by modeling of disk substructures \citep[e.g.,][]{Facchini2018, Alarcon2024}, these gaps are not devoid of material, making extinction a plausible explanation for the non-detections in such systems. 

Theoretical work has also aimed to quantify extinction in disks. \cite{Szulagyi2020} used 3D thermohydrodynamical simulations to explore the effect of dust and gas extinction on the detectability of hydrogen recombination lines in accreting protoplanets. Deriving extinction values under four different opacity assumptions (including both ISM-like and disk-specific dust populations), they concluded that the low detection rate of accreting gas giants is due to planetary emission being strongly absorbed by intervening disk material. \citet{Sanchis2020} used 3D hydrodynamic simulations to model dust distributions and applied standard ISM extinction curves to estimate their effect on protoplanet detectability. They found that for small mass planets of $1~\MJ$ extinction could be as high as $\sim50$ mag at NIR wavelengths. More recently, \citet{Alarcon2024} combined hydrodynamic and radiative transfer models to predict extinction as a function of wavelength and disk location without assuming a predefined extinction law. Their results suggest that large grains can flatten the extinction curve, introducing a grey component that increases attenuation even at long wavelengths such as 4–5~$\mu$m.
Taken together, the observational discrepancies across systems and the reliance of theoretical models on largely untested dust opacities and extinction laws underscore the need for empirical extinction measurement that are independent of assumptions about planetary accretion rates, emission models or dust properties.

All these observational and theoretical works highlight that there is a growing recognition that extinction from disk material may play a crucial role in the non-detection of protoplanets and in the correct interpretation of detected signals, even in disk regions characterized by deep gaps. Furthermore, understanding this effect is vital for optimizing future searches with current and upcoming instruments.

Among the systems with compelling evidence for planet-disk interaction, AS~209 stands out. AS~209 is a K7-type star ($M_* = 1.2\,M_\odot$) at a distance of 121.2~pc \citep{gaia2022} surrounded by a highly structured disk. The distribution of millimeter- and micron-sized grains has been mapped in detail by ALMA/DSHARP and SPHERE/DARTTS-S \citep{Andrews2018, Avenhaus2018}, while the disk’s chemical and physical properties have been characterized by ALMA/MAPS \citep{Oberg2021, Alarcon2021}. The disk exhibits multiple gaps, including a wide ($\approx78$~au) outer gap located at $\sim$200~au, which is clearly traced in both scattered light \citep{Avenhaus2018} and CO emission \citep{Guzman2018}. This deep gap has been interpreted as the result of planet-disk interaction, and has been the subject of detailed modeling of the gas surface density profile \citep{Zhang2021}. Indeed, \cite{Bae2022} used $^{13}$CO line observations to discover a circumplanetary disk candidate at $\sim205$~au from the central star, embedded within the annular gap seen in SPHERE and ALMA observations. Hereafter, we will call the host of this circumplanetary disk as AS~209b.

\begin{figure*}[ht!]
\centering
\includegraphics[width = 0.8\textwidth]{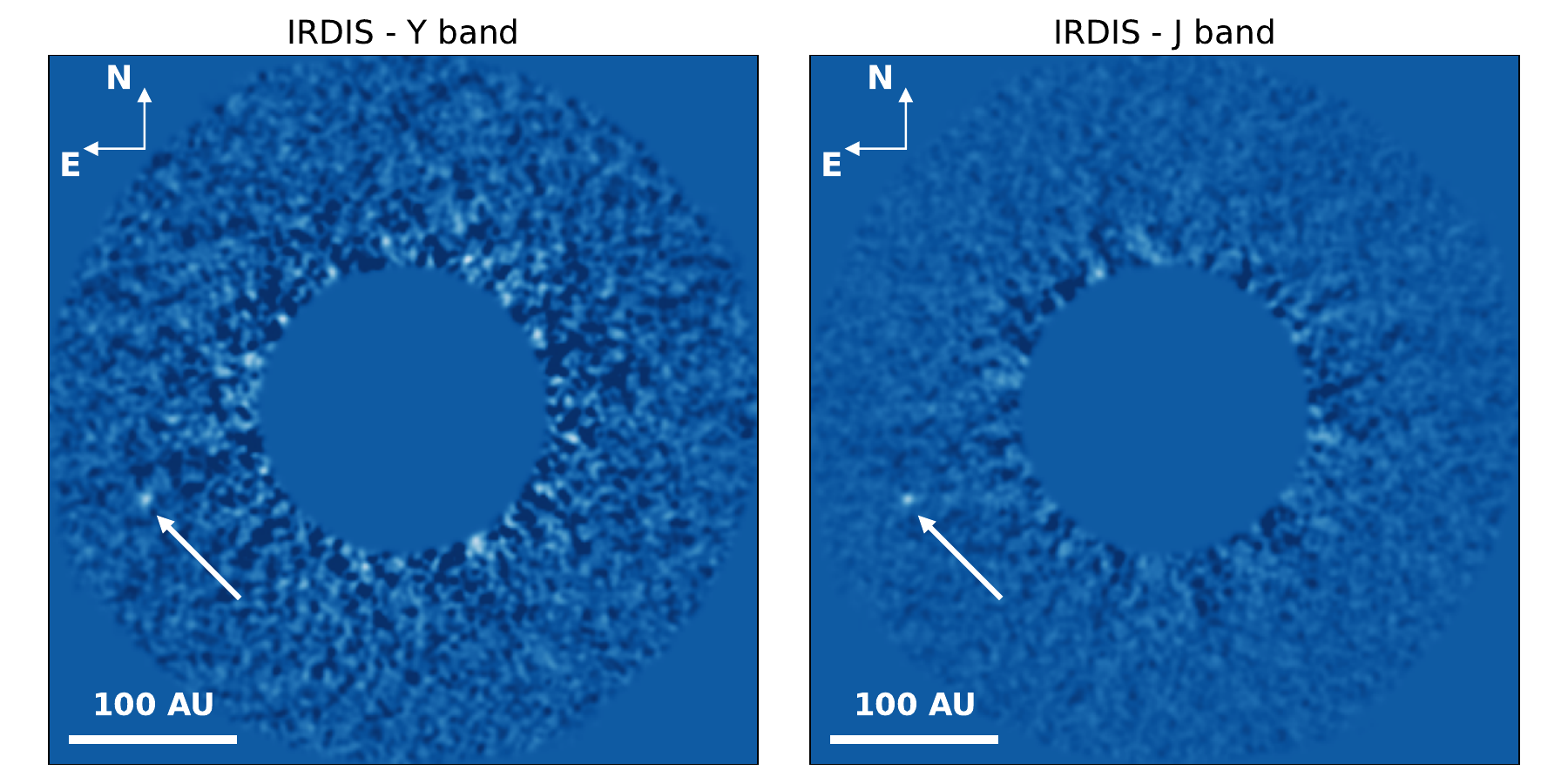}
\caption{$Y$ and $J$ band residuals of AS~209 obtained with VLT/SPHERE. AS~209bkg (CC1) is marked by a white arrow.  
\label{fig:res_sphere}}
\end{figure*}

In this paper, we report the serendipitous discovery of a background star while attempting to image AS~209b, which is coincidentally aligned with the annular gap in which AS~209b is embedded. Archival HST/NICMOS data from 2005 reveal the same object at a similar projected separation, allowing us to classify it as a non-comoving background star. Since that initial detection, we acquired additional data with JWST/NIRCam and VLT/SPHERE, enabling us to build the SED of the object. This fortuitous alignment provides a unique opportunity to directly measure extinction within a planet-opened gap, offering unprecedented insights into the gap material. The paper is structured as follows: Sect.~\ref{sec:observations} and \ref{sec:data_reduction} present the new observations and our data reduction. In Sect.~\ref{sec:results} we build the spectral energy distribution (SED) of the source, which we classify as a background star in Sect.~\ref{sec:classify} using archival HST/NICMOS data from 2005. In Sect.~\ref{sec:spec_fit} we fit the data points to determine which extinction laws are consistent with the data. In Sect.~\ref{sec:discussion} we discuss the results and their implications for planet formation. We conclude in Sect.~\ref{sec:conclusion}.

\section{Observations} \label{sec:observations}

\subsection{VLT/SPHERE}\label{sec:observations_sphere}
We observed AS~209 with the SPHERE/IRDIS \citep{Dohlen2008, Bauzit2019} instrument at the Very Large Telescope (VLT) at Paranal Observatory in Chile. Observations are performed in the $Y$ ($\lambda_c = 1.04~\mu$m) and $J$ ($\lambda_c = 1.24~\mu$m) bands, following the standard procedure for SPHERE coronagraphic imaging. The sequence began with FLUX frames taken using the ND\_N\_1.0 neutral density filter. The star is then placed behind the coronagraph, and satellite spots are activated to obtain CENTER frames, which are essential for accurately locating the stellar position behind the coronagraph. SCIENCE images are subsequently acquired, followed by a second set of CENTER and FLUX frames to complete the observation. All observations were conducted in pupil-tracking mode to enable PSF subtraction using Angular Differential Imaging \citep[ADI;][]{Marois2006}. We had stable conditions during both epochs, although the $Y$-band data were obtained under slightly worse conditions, as indicated by the higher seeing values. The main observational parameters are summarized in Table~\ref{tab:observations_sphere}.

\subsection{JWST/NIRCam coronography}\label{sec:observations_nircam_coro}
We observed AS~209 with JWST/NIRCam using four filters and the round M335R coronagraphic mask (PID~5816, PI: Cugno). For each filter, observations were obtained at two different roll angles to enable ADI. In addition, we observed the reference star HD153135 to perform Reference Differential Imaging (RDI), minimizing self-subtraction and preserving potential extended emission from the disk in the final residuals. HD153135 is observed using the 5-POINT diamond small-grid dither pattern, which samples different stellar positions behind the coronagraph to better characterize and model the PSF. The main observational parameters are provided in Table~\ref{tab:observations_nircam}.

\subsection{JWST/NIRCam imaging}\label{sec:observations_nircam_ima}
Due to the limited set of filters available for coronagraphic observations with NIRCam, we also acquired direct imaging data during JWST Cycles 1 and 3 (PID~2487, PI: Facchini; and PID~5816, PI: Cugno). In Cycle~1, we used the SUB160P subarray to mitigate saturation effects and observed in the F200W and F410M filters. This program aimed to detect the planet candidate associated with the $^{13}$CO signal presented in \cite{Bae2022}, and its results concerning the planet candidate will be presented in a forthcoming paper (Facchini et al., in prep.). The observing strategy followed that of the Cycle~1 program PID~1179 (see, e.g., \citealt{Cugno2024}). In Cycle~3, the SUB320 subarray is used. This configuration resulted in both AS~209 and its disk falling on the stray-light mask between the four detectors of the short wavelength (SW) channel. Consequently, only long-wavelength (LW) observations are included in this work. In the LW channel, AS~209 is observed with the F250M, F323N, and F405N filters.

All imaging data were taken using the RAPID readout mode, with a total of 10 groups per integration to minimize detector artifacts. A 4-point dither pattern is employed to correct for bad pixels and mitigate other detector effects. As with the coronagraphic observations, data were obtained at two roll angles to enable PSF subtraction via ADI. The main parameters of the observations are provided in Table~\ref{tab:observations_nircam}.

\begin{figure*}[ht!]
\centering
\includegraphics[width = 0.8\textwidth]{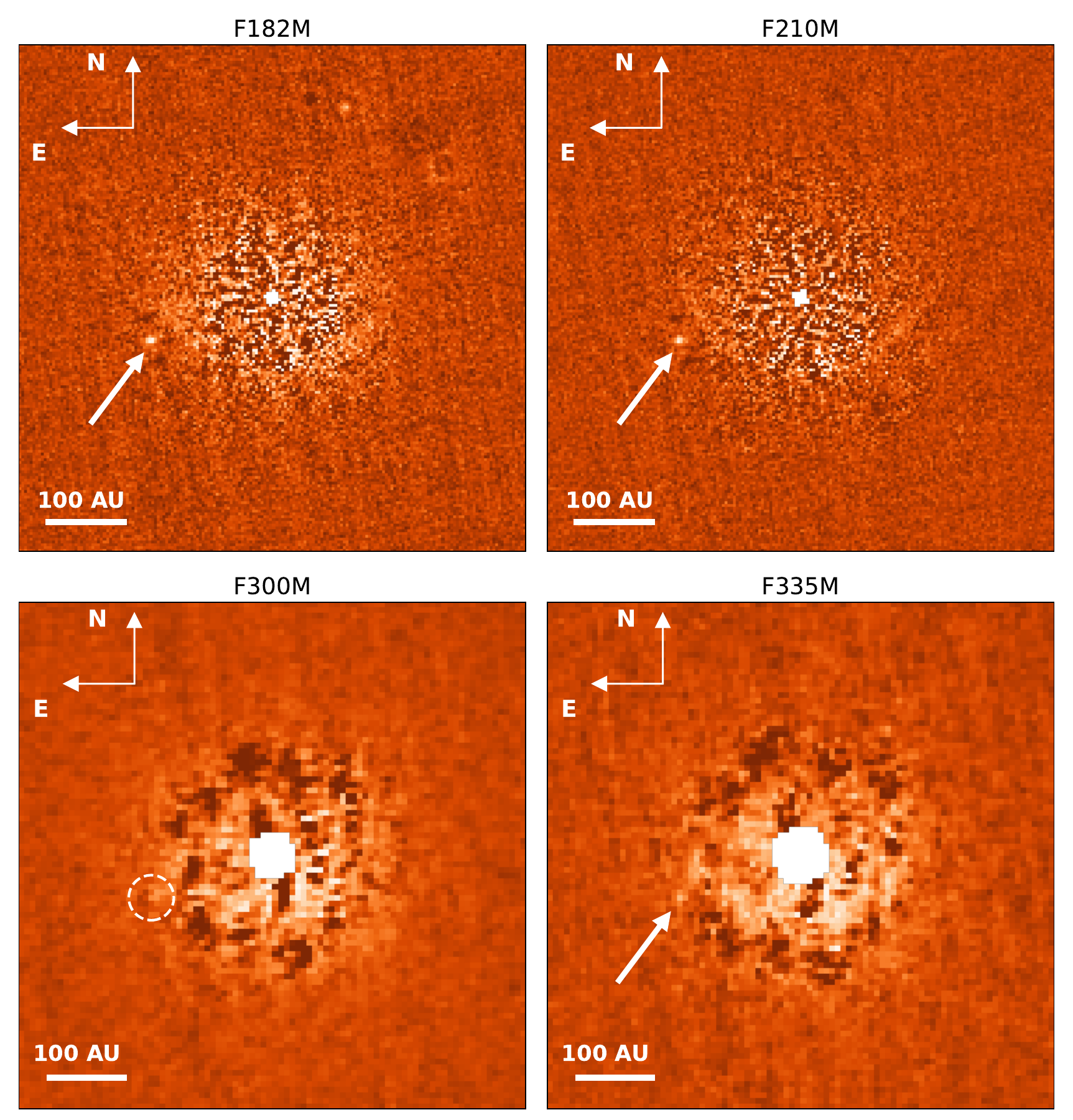}
\caption{JWST/NIRCam coronography residuals for F182M, F210M, F300M and F335M. AS~209bkg (CC1) is marked by a white arrow, while for the non-detection in F300M the position of the source is marked with a white circle.\label{fig:res_nircam_coro}}
\end{figure*}

\section{Data Reduction} \label{sec:data_reduction}

\subsection{VLT/SPHERE/IRDIS} \label{sec:data_reduction_sphere}
We reduced the SPHERE/IRDIS data using the {\tt vlt-sphere} package \citep{Vigan2020}, which performs standard calibrations including dark subtraction, flat-field correction, bad pixel correction, and frame centering while accounting for the instrument's anamorphic distortion. The pre-processed frames are then processed with the {\tt PynPoint} pipeline \citep{Stolker2019}, which carries out PSF subtraction using Principal Component Analysis \citep[PCA;][]{AmaraQuanz2012, Soummer2012}. The resulting residuals are median combined and subsequently convolved with a Gaussian kernel whose full width at half maximum (FWHM) matches that of the instrument PSF.

\begin{figure*}[ht!]
\centering
\includegraphics[width = 0.9\textwidth]{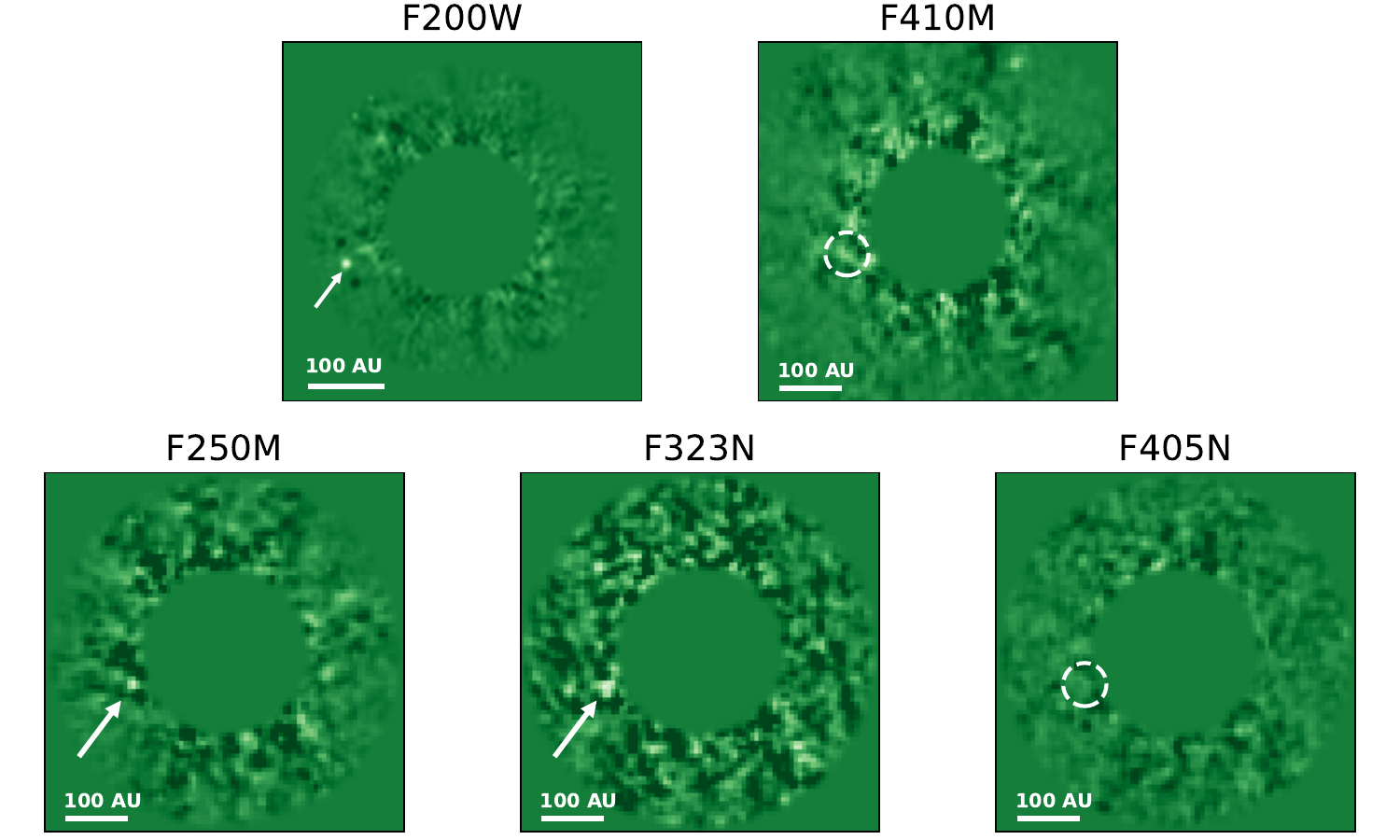}
\caption{JWST/NIRCam imaging residuals for F200W, F410M (top row, Cycle 1) and F250M and F323N and F405N (bottom row, Cycle 3). AS~209bkg (CC1) is marked by a white arrow. North is pointing up in the image, East to the left. The central mask has a radius of $1\farcs1$. \label{fig:res_nircam_ima}}
\end{figure*}

\subsection{JWST/NIRCam coronography}\label{sec:data_reduction_nircam_coro}
We reduced and analyzed the NIRCam coronagraphic data using the {\tt spaceKLIP} package \citep{Kammerer2022, Carter2023}. The {\tt *\_uncal.fits} files are downloaded and processed through Stage 1 and Stage 2 of the {\tt jwst} pipeline, following the recommendations in \citet{Carter2023}. Bad pixels are identified using both the data quality array provided by the pipeline and a sigma-clipping algorithm ($\sigma = 4$), applied to pixels within a three-pixel radius of each target pixel. Flagged bad pixels are replaced with the median value of the four surrounding pixels. The stellar position behind the coronagraph is determined by cross-correlating the images with a synthetic PSF generated using {\tt webbpsf} \citep{Perrin2014}. Images from both visits on AS~209 and the reference star are then aligned. PSF subtraction is performed with {\tt pyKLIP}, integrated within {\tt spaceKLIP}. We found that ADI alone left significant residuals near the stellar PSF, while RDI preserved faint extended emission from the disk at the location of the background source. Combining ADI and RDI allowed us to remove both the disk signal and most stellar residuals, yielding the highest contrast. 

\subsection{JWST/NIRCam imaging}\label{sec:data_reduction_nircam_ima}
The imaging data are reduced following the methodology described in \citet{Cugno2024}. Briefly, we downloaded the {\tt *\_uncal} files and processed them using the {\tt jwst} pipeline (version 1.15.1 with CRDS version 12.0.6), producing the {\tt *\_cal} files without applying dark current correction. As expected for a bright target like AS~209, the inner $0\farcs3$ of the PSF core is saturated and thus unusable. We substitute bad pixels using the PSFs obtained at other dither positions, after shifting the star at the same location of the dither we are correcting. Image centering is performed via spatial cross-correlation with a model PSF generated using {\tt webbpsf}. To mitigate 1/$f$ noise, we masked the PSF structure and subtracted from each row the median of the remaining (unmasked) pixels along that row. Images are then median-combined in groups of 10 frames. 

Stellar PSF subtraction is carried out using the two-roll ADI technique with PCA, implemented via {\tt PynPoint}. For each roll angle, the PSF is modeled using images from the other roll and subtracted accordingly. To suppress systematic alignment errors, the PSFs used for modeling are randomly shifted by values drawn from a normal distribution centered at zero with a standard deviation of 0.05 pixels, following the approach in \citet{Cugno2024}. The resulting residuals are derotated and median-combined to produce the final image.

\section{Results} \label{sec:results}
\subsection{VLT/SPHERE}\label{sec:results_sphere}
Figure~\ref{fig:res_sphere} shows the final residuals from both VLT/SPHERE observations. A companion candidate (CC1) is clearly detected in both the $Y$ and $J$ bands. We used the Markov Chain Monte Carlo (MCMC) implementation in {\tt PynPoint} to insert negative artificial PSFs and iteratively explore the parameter space for contrast, separation, and position angle (PA). To estimate the systematic uncertainties, we injected 360 artificial PSFs at the same contrast but distributed uniformly in PA, and then retrieved them. The final uncertainties are computed by adding in quadrature the MCMC-derived and systematic components. A detailed description of this method is provided in \citet{Stolker2020_miracles}. 

For the $Y$ and $J$ band observations, we measured contrasts of $13.82 \pm 0.24$ and $14.40 \pm 0.10$ mag, respectively. Using stellar flux densities of $0.26 \pm 0.07$~Jy (PAN-STARRS; \citealt{Chambers2016}, based on the mean and standard deviation of two available measurements) and $0.77 \pm 0.03$~Jy (2MASS; \citealt{Cutri2003}) for AS~209, we derive flux densities for CC1 of $0.77 \pm 0.30$ and $1.34 \pm 0.13~\mu$Jy in the $Y$ and $J$ bands, respectively. The larger uncertainty in the $Y$ band measurement is due to slightly worse observing conditions and lower S/N. Additionally, short-wavelength flux measurements of AS~209 show larger scatter, which is reflected in the propagated uncertainty for CC1. Photometric and astrometric results are summarized in Table~\ref{tab:all_points}.

\subsection{JWST/NIRCam Coronagraphy} \label{sec:results_nircam_coro}

Figure~\ref{fig:res_nircam_coro} shows the residuals from the four NIRCam coronagraphic datasets. CC1 is robustly detected in the F182M and F210M filters, while no detection is achieved in F300M. Interestingly, we recover a low-S/N detection in F335M. The absence of a detection in F300M may indicate the presence of water ice in the disk gap material, as ices absorb strongly at $\sim3.0~\mu$m \citep[e.g., ][]{McClure2023, Ballering2025}. 

Astrometry and photometry of CC1 in the F182M, F210M, and F335M filters are extracted using the {\tt spaceKLIP} routines. In addition, we used {\tt spaceKLIP} to estimate the $3\sigma$ detection limits in the F300M filter. These limits are not used in the SED fits (Sect.~\ref{sec:spec_fit}) as the ice feature covered by the F300M filter cannot be properly modeled due to the poor spectral sampling and S/N around $3.0~\mu$m. This threshold is chosen instead of the canonical $5\sigma$, as the location of CC1 is precisely known, allowing for the identification of even marginal signals at that position. All results are summarized in Table~\ref{tab:all_points}.

\subsection{JWST/NIRCam Imaging} \label{sec:results_nircam_ima}

Figure~\ref{fig:res_nircam_ima} shows the residuals from the NIRCam imaging datasets. CC1 is detected in the F200W, F250M and F323N filters, while no detections are achieved in F405N, and F410M. In the non-detection images, the expected position of CC1 is marked with a white dashed circle.

To extract photometry and astrometry of CC1, we used the algorithm described in \citet{Cugno2024}, which is based on MCMC sampling and injection-retrieval of negative point sources, as already applied to the SPHERE data (Sect.~\ref{sec:results_sphere}). Similar to Sect.~\ref{sec:results_sphere}, systematic uncertainties are estimated by injecting fake sources in the data, and are added in quadrature to the MCMC uncertainties. The results for the photometric and astrometric measurements are listed in Table~\ref{tab:all_points}. 

For the F405N and F410M filters, we estimated contrast limits following the method of \citet{Cugno2024}. Using the {\tt applefy} package \citep{Bonse2023}, artificial sources are injected in radial steps of 1 PSF FWHM at six different position angles. A detection threshold of 0.998650 is adopted, corresponding to a $3\sigma$ level under the assumption of Gaussian noise. 

For the property extraction in F200W, F250M and F323N and the contrast limit estimation in other filters, we used P330-E data \citep[see][]{Cugno2024} as a PSF model for photometric calibration. P330-E is a standard G-type star observed during the NIRCam calibration program PID~1538 on 2022 August 29 across all available filters and subarray configurations, and provides a photometrically reliable, unsaturated PSF. The results of all analyses are presented in Table~\ref{tab:all_points}.

\begin{table*}[t!]
\centering
\caption{Summary of the datapoints and limits for AS~209bkg.}
\def\arraystretch{1.25}
\begin{tabular}{lllllllllllll}\hline
Filter              & Flux Star (Jy) & Contrast (mag)   & Flux ($\mu$Jy)    & RA offset (mas) & DEC offset (mas) \\\hline
SPHERE $Y$          & $0.26\pm0.07$   & $13.82\pm0.24$  & $0.77\pm 0.30$   & $1352\pm9$      & $-419\pm14$  \\
SPHERE $J$          & $0.77\pm0.03$   & $14.40\pm0.10$  & $1.34\pm0.13$    & $1347\pm3$      & $-421\pm12$  \\
NICMOS F110W$^\dagger$    & $0.73$    & $11.30\pm0.06$  & $22.31\pm1.17$   & $1215\pm38$     & $-829\pm38$  \\
NIRCam F182M        & $1.33$          & $14.36\pm0.06$  & $2.34\pm0.12$    & $1325\pm3$      & $-460\pm2$    \\
NIRCam F200W        & 0.0201$^*$      & 9.81$\pm0.14$   & $2.39\pm0.32$    & $1342\pm4$      & $-498\pm5$    \\ 
NIRCam F210M        & $1.19$          & $14.18\pm0.06$  & $2.32\pm0.12$    & $1324\pm3$      & $-459\pm2$   \\ 
NIRCam F250M        & 0.01369$^*$     & $9.55\pm0.49$   & $2.16\pm0.75$    & $1335\pm35$     & $-461\pm39$   \\
NIRCam F300M        & $0.69$          & $<13.94$        & $<2.26$          & $-$             & $-$   \\
NIRCam F323N        & 0.00881$^*$     &   $9.27\pm0.76$ & $1.72\pm1.16$    & $1376\pm121$    & $-465\pm151$\\
NIRCam F335M        & $0.65$          & $14.05\pm0.23$  & $1.55\pm0.33$    & $1320\pm18$     & $-463\pm18$   \\
NIRCam F405N        & 0.00584$^*$     & $<8.24$         & $<3.0$           & $-$             & $-$     \\
NIRCam F410M        & 0.00581$^*$     & $<8.01$         & $<3.66$          & $-$             & $-$    \\\hline

\end{tabular}\\\vspace{0.2cm}
\tablenotetext{*}{This flux refers to P330-E, used to calibrate the saturated imaging data.}
\tablenotetext{^\dagger}{Taken from the ALICE survey, \cite{Hagan2018}.}
\label{tab:all_points}
\end{table*}

\section{Classifying the source}\label{sec:classify}

\subsection{A background source shining through the AS~209 disk gap}\label{sec:backgorund_source}
\begin{figure}
    \centering
    \includegraphics[width=0.8\linewidth]{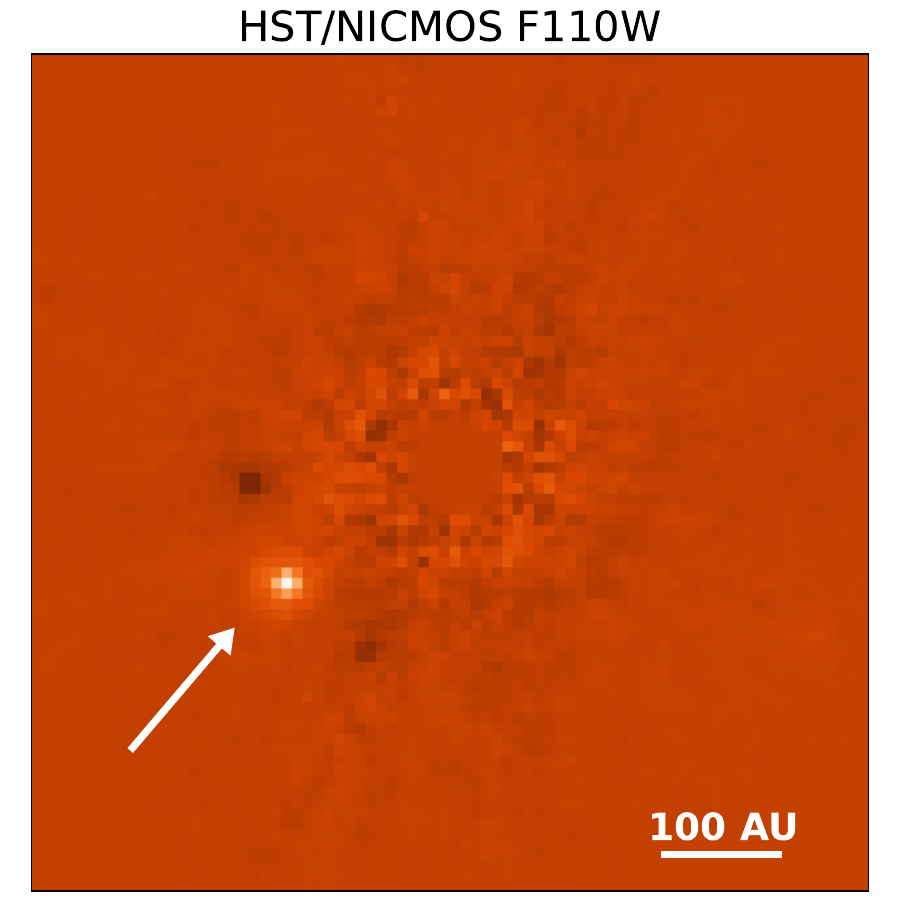}
    \caption{HST/NICMOS F110W residuals obtained from the ALICE survey \citep{Hagan2018}. The white arrow points to AS~209bkg (CC1).}
    \label{fig:F110W}
\end{figure}

To assess whether the detected point source is co-moving with AS~209, we performed a proper motion analysis using an ancillary detection from the ALICE reduction of HST/NICMOS F110W ($\lambda = 1.1~\mu$m) data \citep{Hagan2018}. The archival data, taken on 2005-05-07, show CC1 at a similar separation but with a different PA (see Fig.~\ref{fig:F110W}). Astrometry and photometry from this epoch are retrieved directly from the ALICE database\footnote{\url{https://archive.stsci.edu/prepds/alice/\#dataformat}} and are reported in Table~\ref{tab:all_points}.

\begin{figure*}[ht!]
\includegraphics[width = 0.49\textwidth]{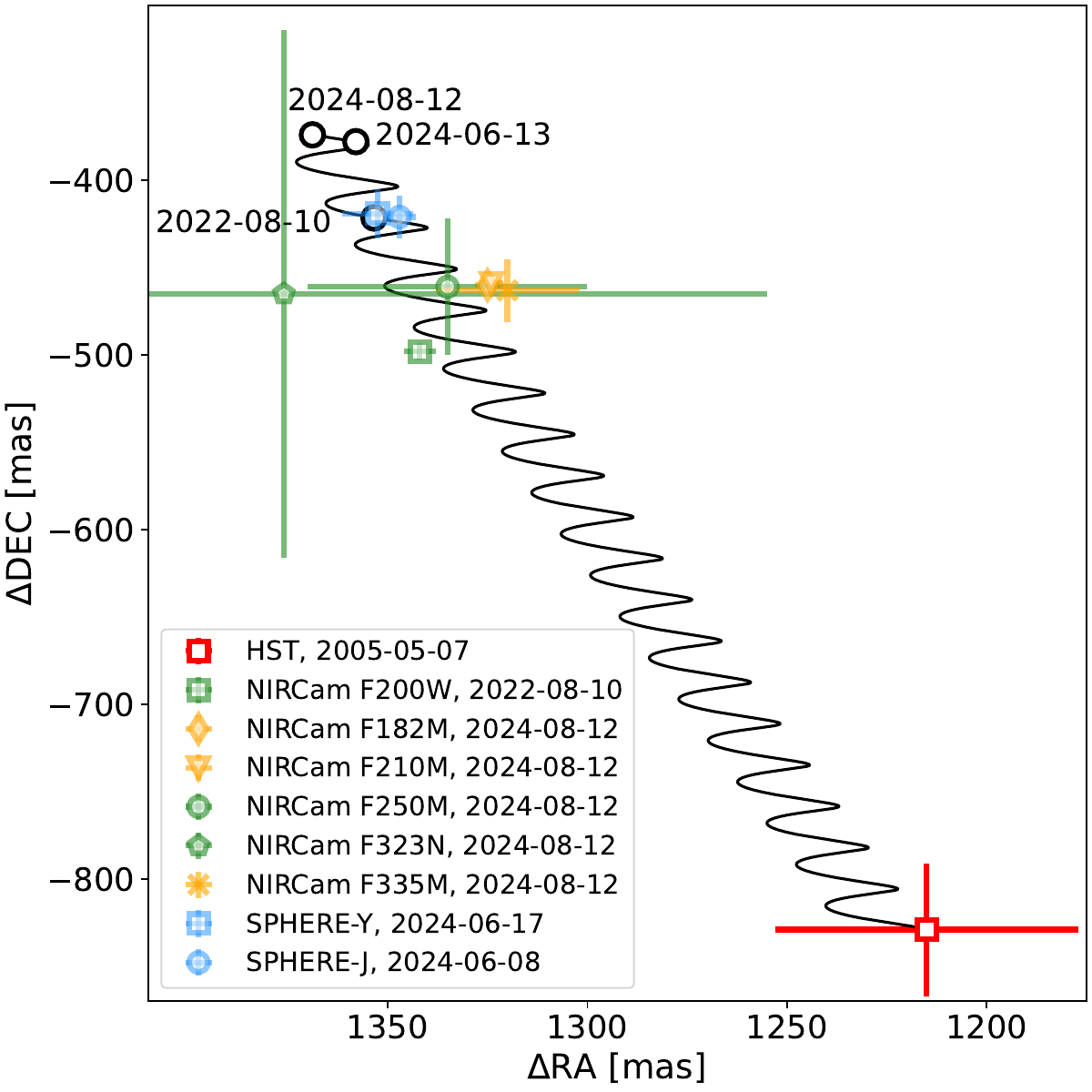}
\includegraphics[width = 0.49\textwidth]{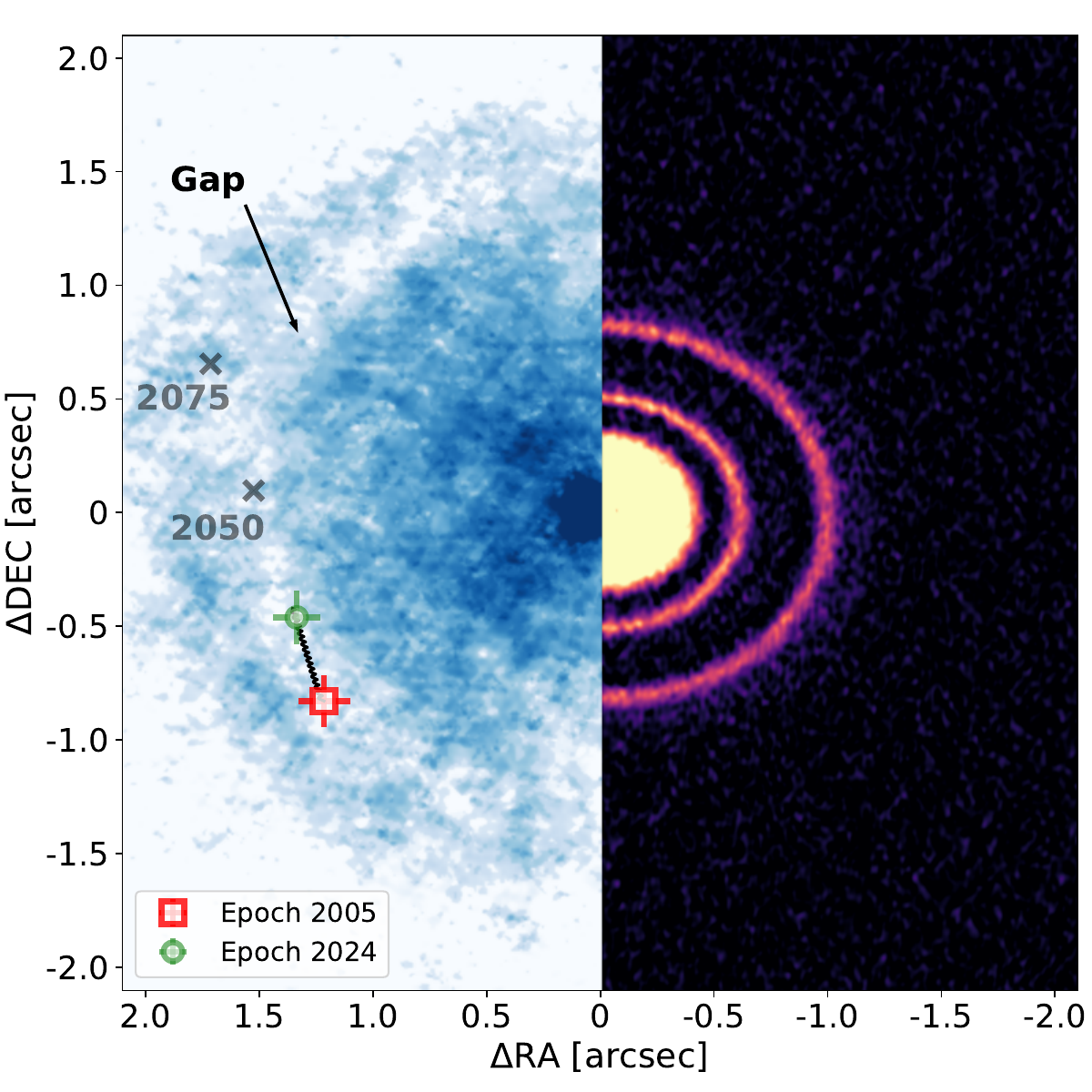}
\caption{Astrometry of the AS~209 system. {\it Left:} Proper motion analysis of AS~209bkg. The red square represent the location of the source in 2005. The black line shows the expected trajectory from 2005 through 2024 assuming it is a stationary object, with the expected locations at the SPHERE and NIRCam epochs highlighted with black circles. The NIRCam and SPHERE data agree with the prediction (when considering the HST uncertainties), demonstrating that the source is indeed a background object. {\it Right:} Projected location of AS~209bkg in the AS~209 disk. On the right side, the millimeter image is reported, showcasing a series of dust rings \citep{Andrews2018}. On the left side, the CO gas as measured by the DSHARP program \citep{Andrews2018, Guzman2018} is reported. AS~209bkg shines through the deep outer CO gap. Grey crosses mark the expected position of AS~209bkg in 2050 and 2075. \label{fig:proper_motion}}
\end{figure*}

We used this 2005 detection to test for common proper motion between AS~209 and CC1. In Fig.~\ref{fig:proper_motion}, we plot all astrometric measurements from this work alongside the 2005 HST detection, color-coded by instrument and observing mode. The black line indicates the expected relative motion of a stationary background object, assuming AS~209 remains fixed, and anchored to the HST detection. Considering the 2005 errorbars, the current location of CC1 is fully consistent with this expected trajectory. We therefore conclude that CC1 is not comoving with AS~209 and is instead a background object. From this point on, we refer to the source as AS~209bkg.

The right panel of Fig.~\ref{fig:proper_motion} shows the projected position of AS~209bkg in the 2024 epoch, revealing that it lies directly behind the deep disk gap identified by \citet{Guzman2018} in CO emission and by \citet{Avenhaus2018} in scattered light. This geometry enables transmission spectroscopy of the AS~209 disk material, using the SED of AS~209bkg as a probe. If the intrinsic properties of AS~209bkg are known, its observed SED can be modeled, and any discrepancy can be attributed to extinction along the line of sight. This extinction includes two components: interstellar extinction between us and AS~209bkg, and extinction from dust within the AS~209 disk gap.

\subsection{What is AS~209bkg?}\label{sec:galactic}
Before studying how dust in the AS~209 disk gap affects the emission of AS~209bkg, it is essential to determine the nature of this background source. As a point-like object, AS~209bkg could be either a Galactic star or an extragalactic object, potentially dominated by emission from an active galactic nucleus (AGN).

\begin{figure}[ht!]
\centering
\includegraphics[width = 0.4\textwidth]{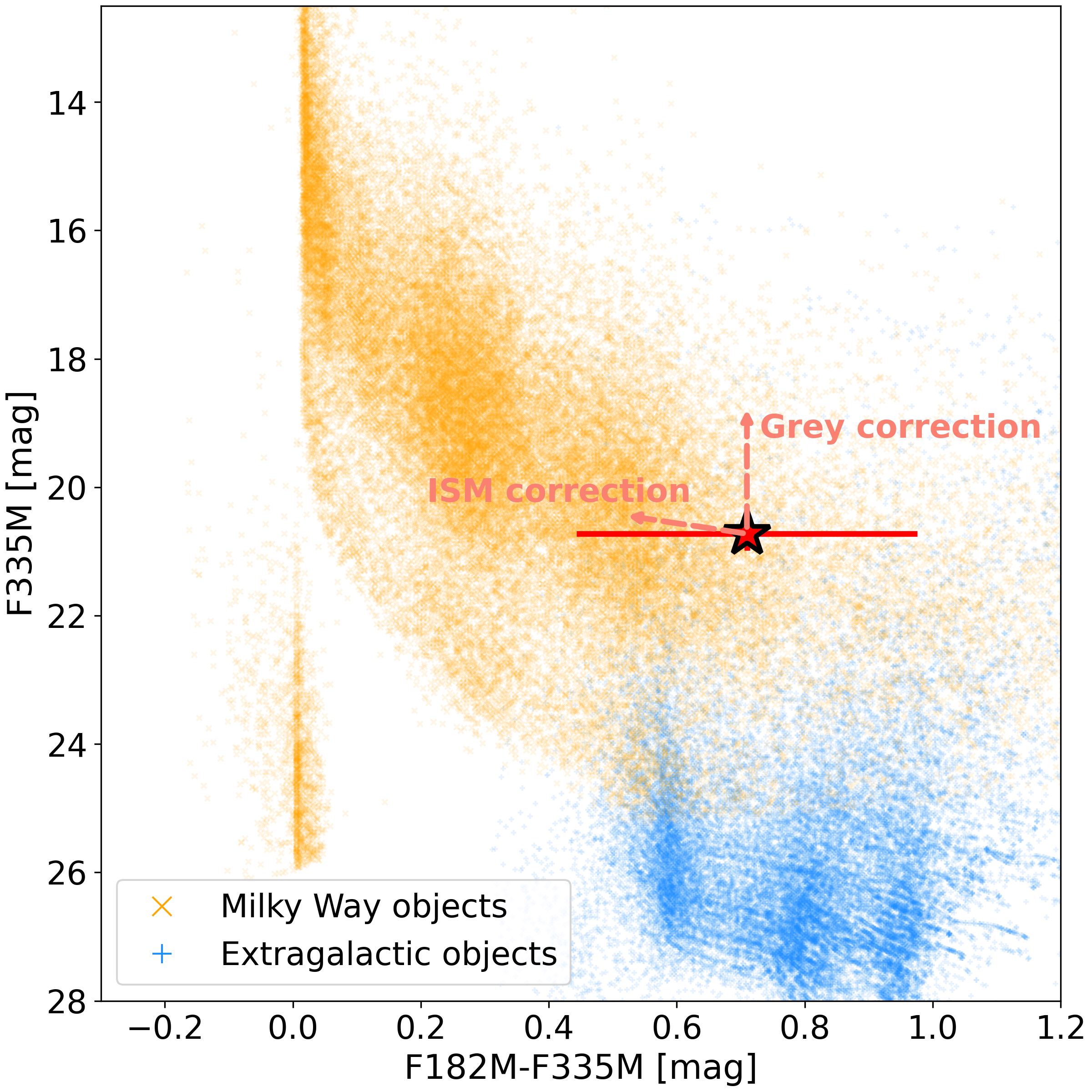}
\caption{CMD of galactic (orange) and extragalactic (blue) objects, and comparison with AS~209bkg. The object is consistent with the galactic population. The arrows show how AS~209bkg would move on the diagram in the case one would correct for ISM or grey extinction ($A_V=2.0$~mag for both). When extinction is accounted for, the photometry of AS~209bkg agrees more strongly with the galactic population.  \label{fig:galactic}}
\end{figure}

To assess this, we compared AS~209bkg’s photometry with synthetic populations of Galactic and extragalactic objects using a color–magnitude diagram (CMD). For the Galactic population, we used the TRILEGAL model \citep{Girardi2012}, which simulates the stellar content along a given line of sight, in this case the coordinates of AS~209, over a 1~deg$^2$ field. For the extragalactic population, we used the JAGUAR catalog \citep{Williams2018}, developed in preparation for the JADES survey. In our CMD, we focus on the F182M and F335M filters, though similar results are obtained with other filter combinations.
In the left panel of Fig.~\ref{fig:galactic}, we show the JAGUAR (blue) and TRILEGAL (orange) sources, and over-plot AS~209bkg as a red star. We note that the magnitudes of AS~209bkg are not de-reddened; extinction vectors are shown for both ISM-like and grey extinction to illustrate the effect of increasing attenuation. By eye, AS~209bkg appears more consistent with the Galactic population, particularly since some extinction (at least ISM) is definitely present along the line of sight.

We followed the method described in \cite{Cugno2024} to quantify the probability that AS~209bkg belongs to either one of the two populations. For each entry in the JAGUAR and TRILEGAL catalogs, we compute a weight based on its proximity in three-dimensional flux space to AS~209bkg’s measured fluxes in F182M, F210M, and F335M. This is done by evaluating a multivariate Gaussian centered on the AS~209bkg fluxes, using the corresponding photometric uncertainties as the Gaussian widths. These weights are summed for each catalog and then rescaled to the relevant area of our observations (400 arcsec$^2$ for the LW NIRCam field of view). We find that the expected number of extragalactic sources consistent with AS~209bkg’s photometry is 0.008, while the number of consistent Galactic sources is 5.2. The ratio between these numbers suggests it is $\sim650$ times more likely that AS~209bkg is a Milky Way object, strongly favoring this interpretation. Including the effects of ISM extinction strengthens this conclusion further, as it shifts AS~209bkg even closer to the bulk of the Galactic population and further from the extragalactic sources in the CMD.

Lastly, we aim to evaluate whether we would obtain plausible properties for the source assuming it would indeed be an extragalactic source. To this end, we first use the well-established code EAZY \citep{Brammer2008} to determine a photometric redshift of the source. The best-fitting solution suggests that it could be a galaxy at very low-redshifts with a peak in the posterior probability distribution at $z\sim0.14$. However, due to the limited wavelength coverage by the photometry compared to most extragalactic surveys, particularly below 1 $\mu$m, the distribution of photometric redshifts is broad, allowing for a range of possible values. We then run the SED modeling code CIGALE \citep{Boquien2019}, which fits the photometric data points with different galaxy and AGN SED models, finding a best fit when using a purely stellar dominated spectrum with no contribution of an AGN. Including the HST/NICMOS F110W data point in the fits does not significantly alter the results. Furthermore, AS~209bkg appears too compact to be a resolved galaxy at $z\sim0.14$. At that redshift, $1\farcs0$ corresponds to 2.5 kpc, and typical dwarf galaxy sizes are $\sim0.6$~kpc \citep[e.g.,][]{Carlsten2021}, corresponding to $\gtrsim0\farcs25$. This value is inconsistent with the unresolved nature of the AS~209bkg signal in the NIRCam and SPHERE images. Hence, it seems unlikely that AS~209bkg is a low-redshift galaxy without any AGN component, further supporting the CMD-based conclusion that AS~209bkg is not an extragalactic object.

\section{Fitting the SED of AS~209bkg}\label{sec:spec_fit} 

We fit the available photometric data points of the SED of AS~209bkg to test various extinction models. Here we use {\tt species} \citep{Stolker2020_miracles} to obtain a BT-NextGen model \citep{Allard2011} that describes the stellar emission of AS~209bkg. The model depends on $\Teff$, $\log(g)$, and metallicity $Z$. We fixed the metallicity to $Z=0$, as its posterior is consistently unconstrained due to the limited constraining power of photometric data alone. To estimate the stellar radius, we used the empirical relationship between the effective temperature and the stellar radius from \cite{Cassisi2019}.

The distance $D$ to AS~209bkg is unknown and treated as a free parameter. However, we applied prior constraints based on the Milky Way's geometry. Specifically, we adopted the scale heights of the thin and thick Galactic disks from \cite{Everall2021}, who derived them using Gaia EDR3 data ($h_\mathrm{thick} = 693$~pc, $h_\mathrm{thin} = 260$~pc). These scale heights are adjusted by $\sin(l)$, where $l = 19^\circ$ is the Galactic longitude of AS~209. We also used their density ratio $\rho_\mathrm{thick}/\rho_\mathrm{thin} = 0.141$ to rescale the relative contributions of the two components. The priors for all other parameters are uniform.

In addition to extinction caused by the disk material, a distant background object like AS~209bkg also suffers from interstellar extinction. To account for ISM extinction on the AS~209bkg's SED, we adopted a standard ISM extinction law with $R_V=3.1$ to estimate the extinction from $A_V$ values obtained from the 3D maps of Galactic dust extinction based on Gaia EDR3 and 2MASS measurements \citep{Lallement2022}. These maps provide the extinction as a function of distance $D$ for the given coordinates. In each step of the forward modeling, we estimate the foreground ISM extinction corresponding to the AS~209bkg's distance under consideration, and add it to the extinction caused by the disk.

\begin{table*}[t!]
\centering
\caption{Median and $1\sigma$ uncertainties for the AS~209bkg and disk extinction parameters obtained from the fits. Below each parameter, the used prior is reported. $\mathcal{U}[X-Y]$ means uniform distribution between $X$ and $Y$.} The last column reports the Bayesian evidence relative to the `None' model.
\def\arraystretch{1.5}
\begin{tabular}{lccccccc}\hline
Extinction  & $\Teff$ [K]  & $\log(g)$   & $D$ [pc]  & A$_V^\mathrm{ISM}$ [mag] & A$_\mathrm{grey}$ [mag] & & log($\mathcal{Z}$) \\ 
 & $\mathcal{U}[2600-7000]$ & $\mathcal{U}[3.0-5.0]$ & $\dagger$ & $\mathcal{U}[0-50]$ & $\mathcal{U}[0-20]$ & \\\hline
None        & $3371^{+61}_{-77}$   & $3.19^{+0.27}_{-0.14}$ & $4347^{+267}_{-364}$ &  $-$    & $-$ & & 0   \\
ISM         & $3086^{+243}_{-265}$ & $3.97^{+0.69}_{-0.65}$ & $2486^{+795}_{-693}$  & $2.7^{+0.7}_{-0.7}$ & $-$ & & 5.1      \\
Grey         & $3600^{+225}_{-190}$ & $3.18^{+0.27}_{-0.13}$ & $1074^{+418}_{-425}$ & $-$          & $4.7^{+1.3}_{-1.5}$ &  & $-2.3$    \\ 
ISM+grey    & $5653^{+902}_{-1539}$ & $4.06^{+0.64}_{-0.71}$ & $620^{+465}_{-315}$ & $5.5^{+1.2}_{-2.1}$  & $8.1^{+1.8}_{-1.0}$ &  & $5.2$    \\
\hline\hline
Extinction  & $\Teff$ [K]           & $\log(g)$              & $D$ [pc]            & log(N$_\mathrm{dust}$ [g/cm$^2$]) & $q$  & log(a$_\mathrm{min}$ [cm])  & log($\mathcal{Z}$) \\ 
 & $\mathcal{U}[2600-7000]$ & $\mathcal{U}[3.0-5.0]$ & $\dagger$ & $\mathcal{U}[-5-0]$ & $\mathcal{U}[0-5]$ & $\mathcal{U}[-5- -1]$ & \\ \hline
DSHARP & $3065^{+264}_{-290}$   & $3.92^{+0.70}_{-0.64}$ & $2452^{+876}_{-727}$ &  $-4.09^{+0.19}_{-0.17}$  & $2.94^{+1.42}_{-1.74}$ & $-4.77^{+0.23}_{-0.16}$ & $3.1$    \\\hline
\end{tabular}
\tablenotetext{\dagger}{A galactic model based on \cite{Everall2021} is used for the prior of the distance.}
\label{tab:fit_results}
\end{table*}


We fit the photometric data from Table~\ref{tab:all_points} using {\tt pymultinest} \citep{Feroz2009, Buchner2016}. The log-likelihood function is defined as
\begin{equation}
    L = -\frac{1}{2} \Big(\log(2\pi\sigma_i^2) + \sum_i \frac{(D_i-M_i)^2}{\sigma_i^2}\Big)
\label{eq:likelihood}
\end{equation}
where $i$ indexes the measurements, $D_i$ and $\sigma_i$ denote the observed flux and its uncertainty in each filter, and $M_i$ represents the model-predicted flux. We exclude F300M from the likelihood calculations because its bandpass may be significantly affected by the $3.0~\mu$m water-ice feature. The F405N, F410M non-detections are treated as hard upper limits: at each iteration, the algorithm checks that the forward model flux does not exceed the observed limit in these bands. If $M_i$ exceeds the limit, the likelihood is set to a negligibly small value.

\subsection{Disk extinction models}
\label{sec:extinction_models}

We test five extinction models, each characterized by a different number of parameters. The `None' model assumes no extinction from the disk material and only considers the interstellar extinction AS~209bkg suffers from described previously as a function of its distance. The `ISM' model adds extinction from the disk using the standard ISM extinction law \citep[$R_V=3.1$,][]{Cardelli1989}, parameterized by the visual extinction $A_V^\mathrm{ISM}$. The `Grey' model includes a wavelength-independent (grey) extinction component, described by a single parameter $A_g$. The `ISM+grey' model combines both previous components, and is described by two free parameters, $A_V^\mathrm{ISM}$ and $A_g$. Finally, the `DSHARP' model derives extinction using the DSHARP opacities \citep{Birnstiel2018}. In this case, the dust population is described by a minimum grain size $a_\mathrm{min}$ (maximum grain size $a_\mathrm{max}$ is left to the standard value of 1~m)  and a power-law exponent $q$, such that $n(a) \propto a^{-q}$, and the total amount of dust is set by the dust surface density $N_\mathrm{dust}$.

\subsection{Fitting the 2022-2024 epoch}
\label{sec:2022-2024}

\begin{figure*}
    \includegraphics[width=\textwidth]{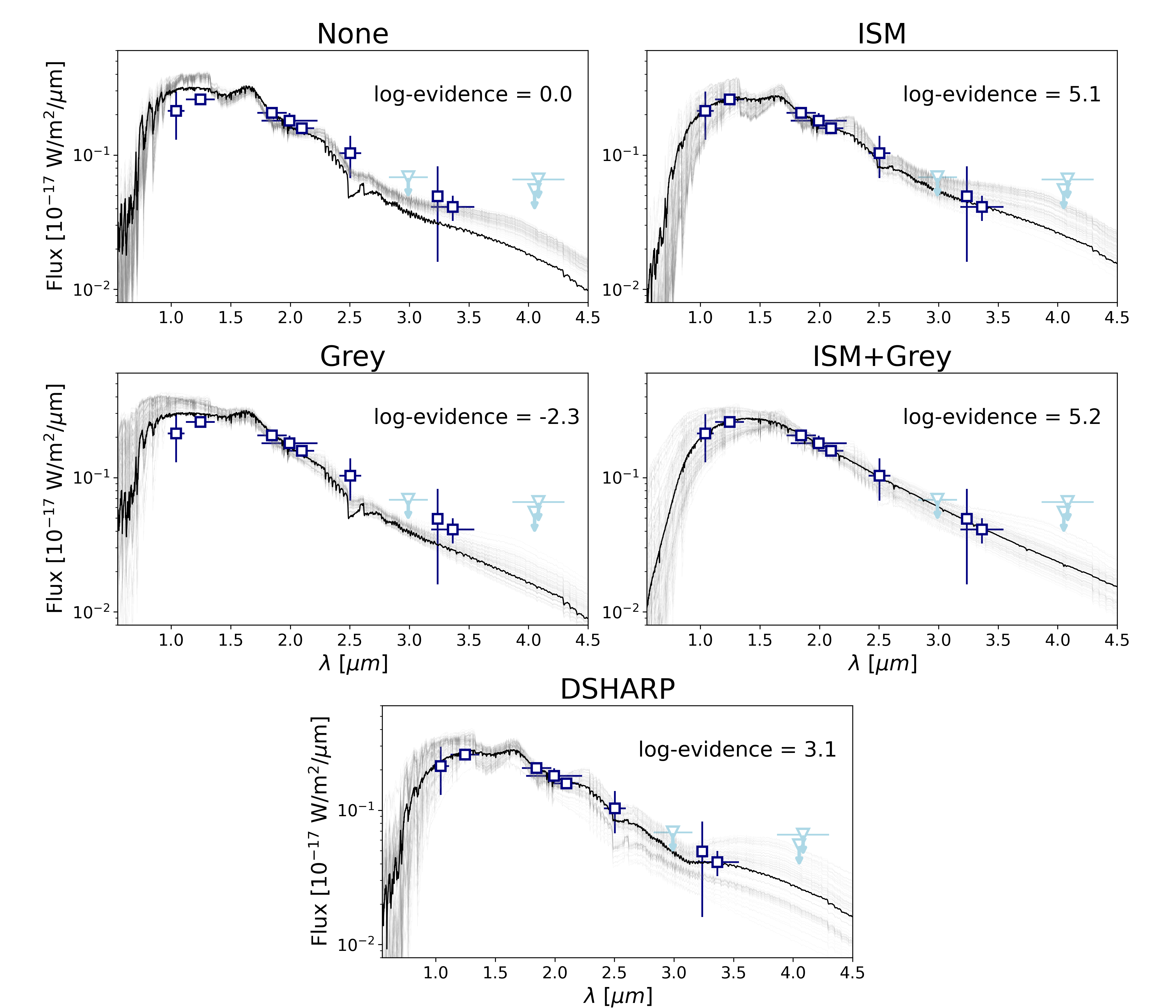}
    \caption{SED fits for AS~209bkg using multiple extinction models. In each panel, the observed fluxes are reported in blue squares, while detection limits are shown in light blue arrows. The F300M detection limits, potentially impacted by the $3.0~\mu$m water-ice band, is shown for completeness but is not used in the fit. The best fit model is shown with a black line, while 100 random samples drawn by the posterior distribution are shown in grey. The log-evidence with respect to the `None' scenario (top left) is reported in each panel.}
    \label{fig:spectra_all}
\end{figure*}

Since in 2005 AS~209bkg was shining through a different region of the disk, we initially focus solely on the SED obtained from the 2022-2024 data. The results of the fits are summarized in Table~\ref{tab:fit_results}. Figure~\ref{fig:spectra_all} shows the 2022--2024 data points (blue), the detection limits (cyan arrows), the best-fit models in black, and a sample of 100 models randomly drawn from the posterior distributions. For three models—`None', `ISM', and `DSHARP'—the best-fit parameters fall outside the $1\sigma$ range of their respective posteriors. This is also evident in Fig.~\ref{fig:spectra_all}, where the spectral shape of the best-fit model diverges from that of the posterior samples. This discrepancy arises because the best-fit model, which is based solely on maximizing the likelihood defined in Eq.~\ref{eq:likelihood}, tends to favor parameter combinations (typically at larger distances) that are disfavored by the priors designed to ensure AS~209bkg resides within the Milky Way.

We use the Bayesian evidence to evaluate the model performance accounting for both priors and likelihood. To ensure that the Bayesian evidence values are meaningfully comparable across models, we adopted priors that are as similar as possible in terms of volume and functional form. Whenever two models share a given parameter (e.g., $\Teff$, $A_V^\mathrm{ISM}$), we used the same prior range and distribution (see Table~\ref{tab:fit_results}). Some models, such as `DSHARP', include parameters that are not present in others, in which case a direct match of prior volumes is not possible. In that case, we verified that the prior-predictive distribution of $A_V$ peaks at low values, with 84\% of the distribution lying below 55 mag, which makes it comparable to the priors for the other models. Bayesian evidence is typically reported relative to a baseline model, which in our case is the `None' model. The final column of Table~\ref{tab:fit_results} lists the relative evidence values for the different extinction models, and these are also indicated in the upper-right corner of each panel in Fig.~\ref{fig:spectra_all}. The models that best explain the data are `ISM' and `ISM+Grey', which yield nearly identical evidence values. `DSHARP' provides an improvement with an evidence of 3.3, even though not as significant as `ISM' and `ISM+Grey'. The remaining models do not offer a statistically significant improvement over the baseline.

These two preferred models point to very different physical scenarios. Indeed, the `ISM' model suggests a relatively cold ($\Teff\sim3100$~K) and further away object ($D\sim2.5$~kpc), while the `ISM+Grey' model suggests a hotter ($\Teff\sim5650$~K) and closer ($D\sim0.6$~kpc) object. In terms of extinction, the `ISM' model yields moderate disk extinction, with a visual extinction of $A_V=2.7 \pm 0.7$~mag. At longer wavelengths, such as $\lambda = 4~\mu$m, the extinction drops substantially, to $A_\mathrm{4~\mu m} = 0.12 \pm 0.03$~mag. In contrast, the `ISM+Grey' model, which suffers from significant correlations between $A_V$, $A_g$ and $D$, implies much heavier extinction, predicting a total extinction (ISM and grey components) of $A_V = 13.7^{+2.4}_{-4.0}$~mag and $A_{4\mu\mathrm{m}} = 8.3^{+1.8}_{-2.0}$~mag.

It is important to note that these values represent the total disk extinction along the line of sight, and a protoplanet embedded in the midplane would experience approximately half this amount. Nonetheless, the implications of the two models are markedly different. The `ISM' model supports the interpretation that high-contrast imaging detection limits are trustworthy, especially in the infrared. On the other hand, the `ISM+Grey' model challenges this view, suggesting that significant extinction could obscure protoplanet emission even at longer wavelengths.

\subsection{Adding the 2005 epoch}
\label{sec:2005}

The results presented in the previous section are based solely on flux measurements of AS~209bkg obtained between 2022 and 2024. However, the 2005 HST detection provides an additional constraint that can be used to refine the SED modeling. Under the assumptions that (i) AS~209bkg is not intrinsically variable, and (ii) the foreground extinction unrelated to the AS~209 disk is the same in 2005 as in 2022–2024, we can include the 2005 flux measurement in our fit.

Because AS~209bkg is located behind a different region of the disk in 2005 approximately $42$~au from its current position, we cannot assume the same disk extinction applies. The flux measured with HST in the F110W filter is $22.3 \pm 1.1~\mu$Jy, indicating that the extinction along that line of sight must have been significantly lower.

To incorporate this constraint, we extended the model to include separate extinction parameters for the two epochs. For the `ISM' and `Grey' models, this involved duplicating the disk extinction parameters: one set for the 2022–2024 epoch and another for 2005. In the `ISM+Grey` model, we assumed the ratio between $A_V$ and $A_g$ remained fixed between the epochs, and introduced a new parameter $A_V^{\mathrm{ISM, }2005}$ to describe the extinction in 2005. In the `DSHARP` model, we kept the dust population parameters ($a_\mathrm{min}$ and $q$) fixed across epochs, but allowed the dust surface density to vary, introducing a new parameter $N_\mathrm{dust}^{2005}$.

\begin{figure*}[t!]
    \includegraphics[width=\textwidth]{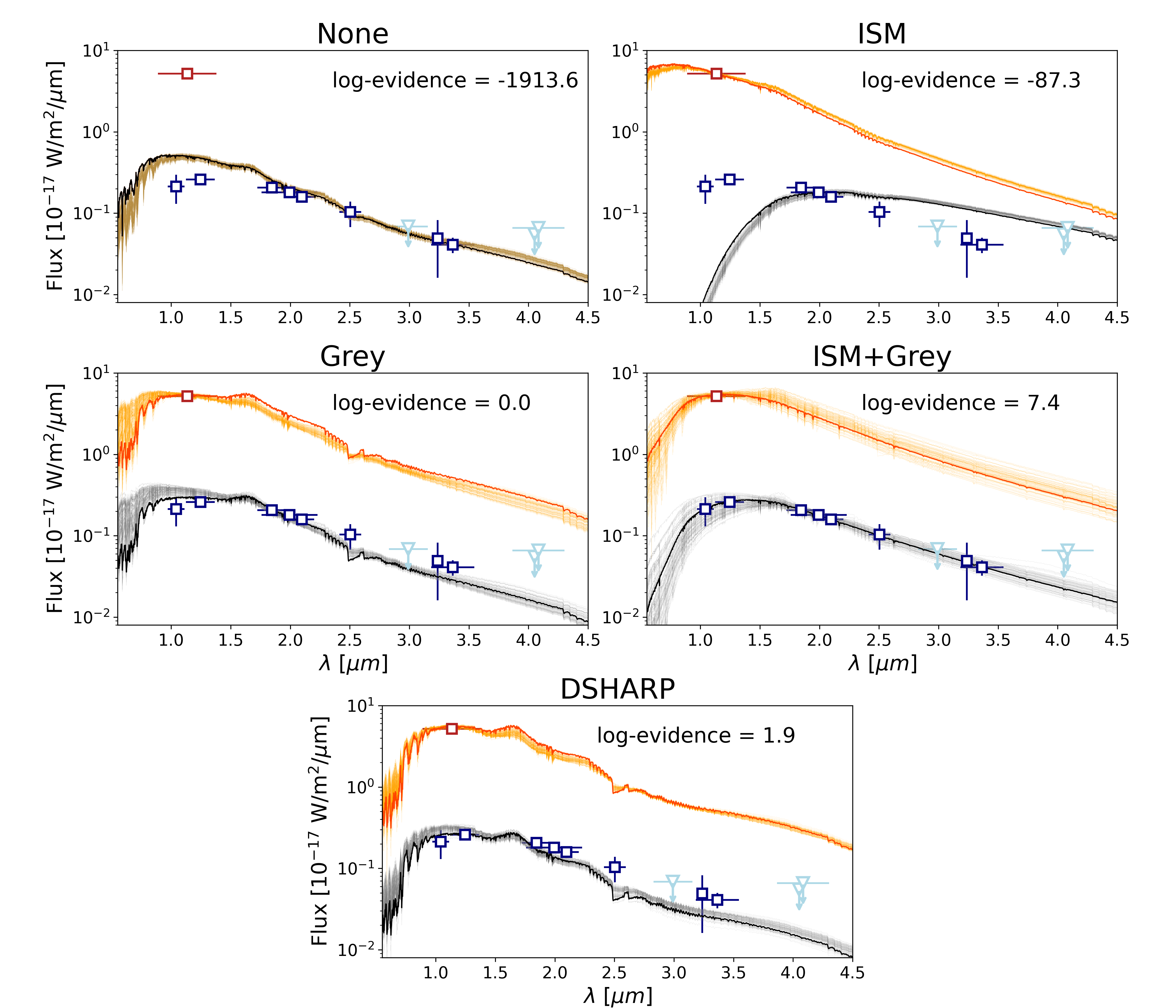}
    \caption{Same as Fig.~\ref{fig:spectra_all}, with the addition of the red square representing the 2005 HST detection, and orange lines that represent the random sample and best-fit model for the 2005 epoch. In this case, the `Grey' model is used to benchmark the Bayesian log-evidences}
    \label{fig:spectra_ep2005}
\end{figure*}

The results of the fits are shown in Fig.~\ref{fig:spectra_ep2005} and reported in Table~\ref{tab:fit_results_2005}. Here, we used the `Grey' model as a reference to compute the difference in log-evidence. This analysis rules out some models that were previously considered viable. In particular, the `ISM` model is now excluded. As reported in Sect.~\ref{sec:2022-2024}, this model implies only modest disk extinction in 2022–2024. Even assuming zero disk extinction in 2005, the unextincted SED predicted by this model is still too faint to match the flux measured by HST. To reconcile this discrepancy, the model instead fits a much hotter source to match the 2005 F110W measurement, requiring substantial extinction to reduce the flux to the level observed in 2022–2024. This leads to an overestimation of extinction at shorter wavelengths ($1.0$–$2.5~\mu$m), inconsistent with the data (see top right panel of Fig.~\ref{fig:spectra_ep2005}). Therefore, under our stated assumptions, extinction consistent with an `ISM'-like law cannot account for the dramatic flux change observed between 2005 and 2022–2024.


The extinction model `ISM+Grey' provides the best fit when the 2005 observation is included (see Fig.~\ref{fig:spectra_ep2005} and Table~\ref{tab:fit_results_2005}). In addition to enforcing a constant $A_V/A_g$ ratio between the two epochs, we explored two alternative scenarios: (i) allowing both $A_V^{\mathrm{ISM, }2005}$ and $A_g^{2005}$ to vary independently, and (ii) fixing $A_V^\mathrm{ISM}$ across both epochs while treating $A_g^{2005}$ as a free parameter. In all cases, the Bayesian evidence indicates that `ISM+Grey' remains the most favored model for explaining the combined dataset. Due to significant degeneracies, the posterior distribution reported in Fig.~\ref{fig:corner} shows a secondary peak at colder effective temperatures ($\sim3000$~K) which implies a significantly lower extinction, with the possible values for $A_V^{\mathrm{ISM, }2005}$ reaching 0~mag and the grey component $A_g$ being as small as $\sim2$~mag. In the next section, we discuss how these degeneracies can be broken.

\section{Discussion}
\label{sec:discussion}

\subsection{Caveats}

Despite the supporting evidence presented in Sect.~\ref{sec:galactic} for the stellar nature of AS~209bkg, the possibility that the source is extragalactic cannot yet be completely ruled out. If AS~209bkg is a background galaxy whose emission is dominated by a central AGN, estimating its extinction would become effectively impossible: without constraints on its distance or intrinsic luminosity, strong degeneracies, particularly with a grey extinction component, would remain unresolved.

The most definitive way to exclude an extragalactic origin is to obtain a spectrum that reveals absorption lines characteristic of stellar atmospheres. In addition, a spectrum with absorption lines can be used to determine the spectral type of AS~209bkg, breaking some of the remaining degeneracies in the fit. However, these observations are challenging due to the extremely low brightness of the object and the high contrast required. An alternative approach would be to measure the parallactic motion of AS~209bkg, thereby confirming its Galactic origin. This would require two-epoch astrometric observations with GRAVITY at the VLTI.

In addition to confirming the Galactic nature of the object, such measurements would yield a much tighter prior on its distance, which could be directly incorporated into the SED fitting. Rather than relying on broad priors that only require AS~209bkg to lie within the Galactic disk, a direct distance measurement would allow for more precise constraints. As shown in Tables~\ref{tab:fit_results} and~\ref{tab:fit_results_2005}, the current $1\sigma$ uncertainties on the posterior distance distributions span a wide range. Reducing this uncertainty would likely further exclude some models and significantly tighten the extinction estimates. In particular, several extinction parameters (such as $A_\mathrm{grey}$) exhibit strong degeneracy with the assumed distance to AS~209bkg.

\subsection{Connecting variability with disk geometry}
\label{sec:disk_geometry}

The fact that the two SPHERE filters overlap in wavelength with the F110W measurement from 2005 confirms that the observed variability is real and cannot be attributed to unusual spectral features. We explore two possibilities that do not involve changes in the AS209 disk extinction to explain some of the measured variability between 2005 and 2022, which amounts to a factor of 22. First, intrinsic stellar variability could account for some of the discrepancy, but such amplitude is rare, especially in the near-infrared. Second, changes to foreground extinction unrelated to the disk could also cause some of the variation, but such a change in flux at $1.1~\mu$m would correspond to an increase in visual extinction of approximately $A_V = 9.7$~mag. Such a gradient in interstellar extinction is unlikely, given that AS~209bkg’s projected motion between 2005 and 2022 is only $\sim0\farcs35$. Hence, external factors seem to be unable to explain the significant flux change.

\begin{figure}
    \centering
    \includegraphics[width=\linewidth]{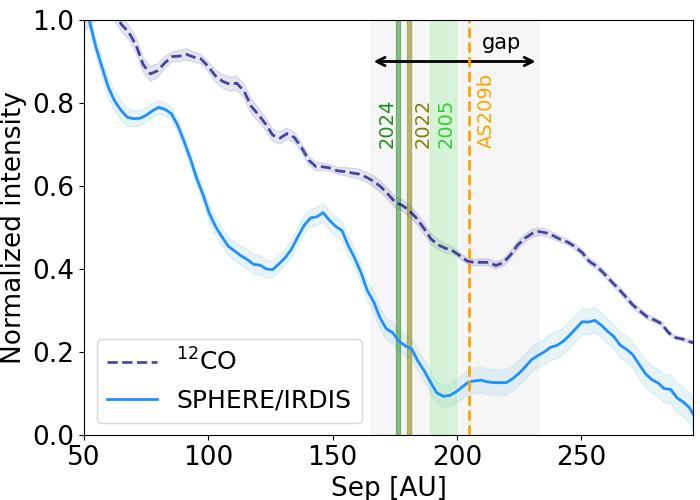}
    \caption{AS~209 disk intensity profile in polarized $H$-band scattered light (solid light blue line, \citealt{Avenhaus2018}) and peak intensity of $^{12}$CO (dashed dark blue line, obtained by azimuthally averaging the unobscured East side of the data from \citealt{Andrews2018,Guzman2018} with the GoFish package, \citealt{Teague_gofish}). Vertical shaded areas report the radial location ($1\sigma$ uncertainties) of AS~209bkg in the three epochs presented in this paper, while the dashed orange line shows the radial separation of AS~209b \citep{Bae2022}. The grey area shows the inner and outer radius of the wide gap observed in the disk \citep{Guzman2018, Avenhaus2018}. }
    \label{fig:profile}
\end{figure}

In contrast, this strong flux variation is qualitatively consistent with the expected structure and geometry of the disk. As shown in Fig.~\ref{fig:proper_motion}, AS~209bkg has been moving steadily closer to the inner edge of the disk gap since 2005, with its deprojected separation from the central star decreasing from $194.5 \pm 5.2$~au in 2005 to $180.7 \pm 0.8$~au in 2022, and $176.4 \pm 0.7$~au in 2024 (we adopted $i = 35.0 \pm 0.1^\circ$ and $\mathrm{PA} = 85.8 \pm 0.2^\circ$ from \citealt{Huang2018} for the disk geometry).

Figure~\ref{fig:profile} shows the $r^2$-scaled scattered light profile of the AS~209 disk (light blue solid line) from \citet{Avenhaus2018}, where the scaling compensates for the radial dilution of stellar irradiation, and the $^{12}$CO intensity profile derived from the data presented in \citet{Andrews2018, Guzman2018}.  Vertical lines indicate the radial positions of AS~209bkg at each epoch, and the vertical orange dashed line marks the location of the AS~209b CPD candidate \citep{Bae2022}. At the gap edges (shaded grey area), the scattered light and CO signals increase, reflecting an increase in the disk’s surface density. Thus, as AS~209bkg approaches the gap edge during the 2022–2024 period, its emission traverses denser regions of the disk, experiencing higher extinction than in 2005.

Continued monitoring of AS~209bkg will enable us to continue tracing the radial variation of the dust optical depth and extinction law along the inner edge of the gap. This presents a unique opportunity to map opacities and surface density through the disk gap as a function of radius using a background object in relative motion. Indeed, the radial profile of extinction presented here is based on only two datapoints, but it can be used to constrain the structure of the inner edge of the disk gap. Combining the results of the extinction to the multi-wavelength scattering properties of dust grains across the disk gap has the potential to provide unique constraints on the absorption and scattering opacities of dust grains in the outer regions of the disk. Future dedicated thermo-chemical models of both the dust properties and high-sensitivity line data \citep[such as as the ones provided by MAPS,][]{Oberg2021} will be able to shed light on physico-chemical changes in the disk properties across the AS~209 outer gap.

\subsection{Implications for AS~209b}
In this section we use the additional information provided by this work to re-evaluate the high-contrast imaging non-detections of the AS~209b CPD candidate proposed by \cite{Bae2022}. Figure~\ref{fig:profile} shows the separation of AS~209b with respect to the radial location of AS~209bkg in the different epochs. The constraints measured for the 2005 epoch seem to be more reliable in estimating the extinction the planet suffers from. 

\cite{Cugno2023_magaox} estimated the H$\alpha$ detection limits for AS~209b from MagAO-X data to be $2.5\times10^{-16}$~erg~s$^{-1}$~cm$^{-2}$. The posterior distribution for the extinction in 2005 from Sect.~\ref{sec:2005} suggests that $A_{H\alpha}=4.2^{+0.9}_{-1.2}$~mag. Here, we divided the value obtained by the posterior by a factor of two, under the assumption that AS~209b is located in the disk midplane and therefore its emission would suffer from only half of the extinction measured for AS~209bkg. This results in an extinction corrected H$\alpha$ flux limit for AS~209b of $1.2^{+1.5}_{-0.8} \times 10^{-14}$~erg~s$^{-1}$~cm$^{-2}$, implying that only $\sim2.0\%$ of the photons are able to escape the disk. Using the shock-heated gas models for planetary mass objects presented in \cite{Aoyama2018, Aoyama2020}, and assuming $M_p=2~M_J$ and $R_p = 2~R_J$, one obtains a mass accretion rate of $<1.2 ^{+2.4}_{-0.8}\times10^{-7}~M_J$~yr$^{-1}$. This value can be compared with the mass accretion rate estimated assuming no extinction is affecting the line emission, $M_\mathrm{acc} = 3.1\times10^{-9}~M_J$~yr$^{-1}$: the actual limits after the correction for extinction are almost 2 orders of magnitude higher. As shown in Fig.~5 of \cite{Cugno2023_magaox}, this value is compatible with the average mass accretion rate necessary to build a gas giant planet as massive as AS~209b in the timeframe of the age of the system ($1-2$~Myr, \citealt{Andrews2009}). It is important to note that this calculations only consider the extinction from the circumstellar disk material and neglects any additional extinction from the circumplanetary environment. Hence, these extinction values should likely be treated as a lower limit. 

In the thermal regime, AS~209 has been observed with VLT/NaCo in the $L'$ band \citep{Jorquera2021, Cugno2023_ISPY}. We consider the upper limits provided in \cite{Cugno2023_ISPY}, who reported an apparent magnitude detection limit of 17.2 mag. To avoid assumptions on planet initial condition, age, and atmospheric and environment properties, the authors used blackbodies to quantify the depth of the observations. The estimated limit without accounting for extinction corresponds to an effective temperature limit of $\sim830$~K when assuming $R_p=2~R_J$. Our best-fit model at the 2005 epoch predicts an extinction of $2.7^{+0.7}_{-0.7}$~mag, implying a true extinction-corrected apparent magnitude limit of 14.5~mag. This new value corresponds to an estimated effective temperature of $\sim1760$~K, implying that a massive planet forming following a hot-start model could still be forming in the disk and not be detected by our instruments. The F410M NIRCam data provide a tighter constraint on the planet thermal flux, and will be presented in a forthcoming paper (Facchini et al., in prep.).

\subsection{Implications for protoplanet searches}

Analysis of multiple datasets of protoplanetary disks at H$\alpha$ wavelengths \citep{Cugno2019, Zurlo2020, Huelamo2022, Follette2023} and in the NIR \citep{AsensioTorres2021, Jorquera2021, Cugno2023_ISPY, Ren2023, Wallack2024}, even with JWST \citep{ Cugno2024, Wagner2024, Mullin2024}, resulted in a series of non-detections. In multiple cases, the authors pointed out that extinction could be one of the causes for these non-detections, but empirical evidence and a direct measurement were still lacking.

The extinction measurement from the material in the gap of the AS~209 disk is an important first measurement that explains and reconciles the ALMA disk observations pointing towards forming gas giant protoplanets interacting with the disk material with the deep non-detections provided by high-contrast imaging searches. This measurement suggests that some level of extinction is to be expected in disks and needs to be accounted for when interpreting non-detections. However, every disk is unique, and the extinction exerted by the disk material on the protoplanet emission depends on the disk dust surface density, the dust population properties, and the gap structure, depth, and geometry. The AS~209 gap is indeed peculiar: it is observed in CO and scattered light, but is located in the disk outer regions where the radio surface brightness is too low to be detected \citep{Bae2022}. Extrapolating the conclusions drawn for this gap to mm gaps orbiting at smaller radial separations is a very difficult task. On one hand, dust closer to the star is expected to present higher surface densities and larger grain sizes due to radial drift, thus increasing the expected extinction. On the other hand, deep gaps or cavities (as the one in PDS~70) may be so devoid of dust to drastically reduce ambient extinction. More studies following the approach presented in this paper can however break these degeneracies across different disks and gap properties.

\section{Conclusions}
\label{sec:conclusion}
This work introduces a novel approach to probe the planet-forming environment and its material, by performing transmission spectroscopy on a background object shining through the AS~209 disk. 
Based on the source proper motion and the color-magnitude diagram we determine that the background source is associated with the galactic population and is not a distant AGN.  With this knowledge, we use the detected SED of the background source and models of the galactic dust attenuation to extract the dust extinction from the dust gap in AS 209. By measuring the extinction across a broad wavelength range, we demonstrate that dust within the disk gap exerts significant attenuation in the optical to near-infrared (NIR), with clear implications for the detectability of embedded protoplanets. These findings help reconcile the tension between strong ALMA-based indicators of ongoing planet-disk interaction and the lack of detections in high-contrast imaging observations.

Our best fit extinction model includes two components: (i) an ISM-like extinction arising from small grains that permeate through the gap and (2) a grey component due to somewhat larger grains. The presence of a grey extinction component in the NIR has particularly important consequences at a time when the community is increasingly focusing on the $4-5~\mu$m range accessible with JWST/NIRCam \citep[e.g.,][]{Cugno2024}. While both simulations and early observations suggest that circumplanetary disks can boost planetary flux in the MIR \citep[$10-15~\mu$m,][]{Szulagyi2019, Choksi2025, Cugno2024_gqlup}, where JWST/MIRI operates, the extent to which the grey extinction component persists at these wavelengths remains uncertain. Additional data are needed to determine whether extinction becomes negligible or remains a limiting factor even at these wavelengths.

Finally, this study motivates regular monitoring of background stars viewed through disks, which can be used to directly map extinction at different radii and azimuthal locations. By combining new observations with archival data, we can reconstruct the dust surface density structure in a range of disks. In addition to shedding light on the local environments where planets form, such studies will place AS~209’s extinction properties in a broader statistical context, helping to determine whether its outermost gap is unusually obscured or representative of planet-hosting disk regions more generally.

\begin{acknowledgments}
We would like to thank the anonymous referee, whose constructive comments improved the quality of this manuscript.
GC thanks the Swiss National Science Foundation for financial support under grant numbers P500PT\_206785 and P5R5PT\_225479. JB acknowledge support from program \#JWST-GO-02487.008-A, which was provided by NASA through a grant from the Space Telescope Science Institute and is operated by the Association of Universities for Research in Astronomy, Inc., under NASA contract NAS 5-03127.SF acknowledges financial contributions by the European Union (ERC, UNVEIL, 101076613), and by PRIN-MUR 2022YP5ACE. MB acknowledges support from the European Research Council (ERC) under the European Union's Horizon 2020 research and innovation program (PROTOPLANETS; grant agreement No. 101002188). Views and opinions are however those of the author(s) only and do not necessarily reflect those of the European Union or the European Research Council. Neither the European Union nor the granting authority can be held responsible for them.

This work is based on observations made with the NASA/ESA/CSA James Webb Space Telescope. The data were obtained from the Mikulski Archive for Space Telescopes at the Space Telescope Science Institute, which is operated by the Association of Universities for Research in Astronomy, Inc., under NASA contract NAS 5-03127 for JWST. These observations are associated with programs GO 2487 (Grant \#JWST-GO-02487.002-A) and GO 5816 (Grant \#JWST-GO-05816.002-A).
The specific observations analyzed in this work can be accessed via \url{http://dx.doi.org/10.17909/tk45-eg60}. This work is based on observations collected at the European Southern Observatory under ESO programme 113.268U.001

This paper makes use of the following ALMA data: ADS/JAO.ALMA\# 2016.1.00484.L, ADS/JAO.ALMA\# 2013.1.00226 and ADS/JAO.ALMA\# 2015.1.00486.S. ALMA is a partnership of ESO (representing its member states), NSF (USA) and NINS (Japan), together with NRC (Canada), NSC and ASIAA (Taiwan), and KASI (Republic of Korea), in cooperation with the Republic of Chile. The Joint ALMA Observatory is operated by ESO, AUI/NRAO, and NAOJ.
\end{acknowledgments}

%

\vspace{5mm}
\facilities{JWST(NIRCam), VLT/SPHERE, HST(NICMOS)}


\software{Pynpoint \citep{Stolker2019}, 
          spaceklip (version 2.7.1) \citep{Kammerer2022, Carter2023},
          species \citep{Stolker2020_miracles}, 
          GoFish \citep{Teague_gofish}, 
          bettermoments \citep{teague_bettermoments}, 
          jwst \citep{jwst_pip}, 
          pymultinest \citep{Feroz2009, Buchner2016}, 
          EAZY \citep{Brammer2008},
          CIGALE \citep{Boquien2019}}



\appendix

\section{Corner plot of the `ISM+grey' extinction model}
\begin{figure*}[ht!]
\includegraphics[width = \textwidth]{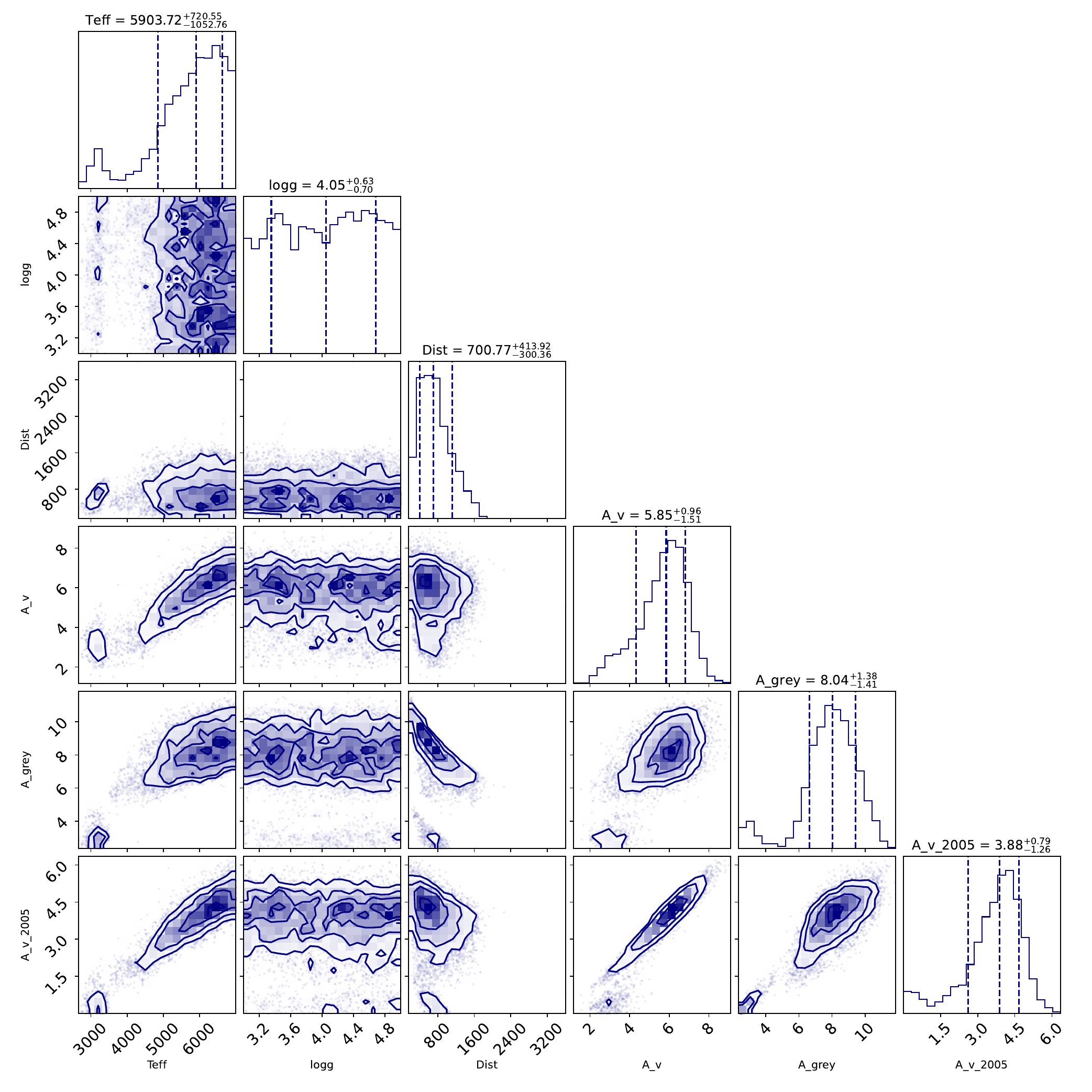}
\caption{Corner plot reporting the posterior distribution for the `ISM+Grey' model.\label{fig:corner}}
\end{figure*}

\startlongtable
\begin{longrotatetable}
\begin{deluxetable}{lccccccccc}
\centering
\tablecaption{Summary of the datapoints and limits for AS~209bkg when including the 2005 epoch. Below each parameter, the used prior is reported. $\mathcal{U}[X-Y]$ means uniform distribution between $X$ and $Y$. The last column reports the Bayesian evidence relative to the `Grey' model}
\def\arraystretch{1.5}
\startdata
Extinction  & $\Teff$ [K]  & $\log(g)$   & $D$ [pc]  & A$_V^\mathrm{ISM}$ [mag] & A$_\mathrm{grey}$ [mag] & A$_V^{\mathrm{ISM}, 2005}$ [mag]  &  A$_\mathrm{grey}^{2005}$ [mag] &  log($\mathcal{Z}$) \\ 
 & $\mathcal{U}[2600-7000]$ & $\mathcal{U}[3.0-5.0]$ & $\dagger$ & $\mathcal{U}[0-50]$ & $\mathcal{U}[0-20]$ & $\mathcal{U}[0-50]$ & $\mathcal{U}[0-20]$ & \\\hline
None        & $3445^{+155}_{-73}$   & $4.96^{+0.03}_{-0.07}$ & $3837^{+3827}_{-270}$ &  $-$    & $-$ & &   & -1913.6   \\
ISM         & $5047^{+51}_{-71}$ & $3.87^{+0.70}_{-0.60}$ & $9437^{+294}_{-436}$  & $18.8^{+0.4}_{-0.3}$ & $-$ & $0.04^{+0.05}_{-0.03}$ &  &  -87.3     \\
grey         & $3612^{+244}_{-140}$ & $3.18^{+0.27}_{-0.13}$ & $946^{+329}_{-349}$ & $-$          & $5.0^{+1.1}_{-0.9}$ & & $2.0^{+1.1}_{-0.9}$ &  $0.0$    \\ 
ISM+grey    & $5903^{+721}_{-1053}$ & $4.05^{+0.63}_{-0.70}$ & $701^{+413}_{-300}$ & $5.9^{+1.0}_{-1.5}$  & $8.0^{+1.4}_{-1.4}$ & $3.9^{+0.8}_{-1.3}$ &  &  $7.4$    \\\hline\hline
Extinction  & $\Teff$ [K]           & $\log(g)$              & $D$ [pc]            & log(N$_\mathrm{dust}$ [g/cm$^2$]) & $q$  & log(a$_\mathrm{min}$ [cm])  & log(N$_\mathrm{dust}^{2005}$ [g/cm$^2$]) &  log($\mathcal{Z}$) \\ 
 & $\mathcal{U}[2600-7000]$ & $\mathcal{U}[3.0-5.0]$ & $\dagger$ & $\mathcal{U}[-5-0]$ & $\mathcal{U}[0-5]$ & $\mathcal{U}[-5- -1]$ & $\mathcal{U}[-5-0]$ \\ \hline
DSHARP & $3390^{+51}_{-61}$   & $3.17^{+0.23}_{-0.13}$ & $1074^{+58}_{-63}$ &  $-0.42^{+0.29}_{-0.45}$  & $2.98^{+1.37}_{-1.59}$ & $-1.48^{+0.31}_{-0.45}$ & $-3.61^{+1.12}_{-0.93}$ & $1.9$    \\\hline
\enddata
\tablenotetext{\dagger}{A galactic model based on \cite{Everall2021} is used for the prior of the distance.}

\label{tab:fit_results_2005}
\end{deluxetable}
\end{longrotatetable}


\bibliography{sample631}{}
\bibliographystyle{aasjournal}



\end{document}